\documentstyle[twocolumn,aps,prl,graphicx,floats]{revtex}
\begin{document}
\draft
\flushbottom
\twocolumn[
\hsize\textwidth\columnwidth\hsize\csname@twocolumnfalse\endcsname

\title{Renormalized mean-field theory of the neutron scattering in
       cuprate superconductors}

\author{Jan Brinckmann$^1$ and Patrick A.\ Lee$^2$}
\address{
  $^1$Institut f\"ur Theorie der Kondensierten Materie,
  Universit\"at Karlsruhe, D-76128 Karlsruhe, Germany \\
  $^2$Department of Physics, Massachusetts Institute of Technology, Cambridge,
  Massachusetts 02139}

\widetext

\date{ 6 July 2001 }

\maketitle
\tightenlines
\widetext
\advance\leftskip by 57pt
\advance\rightskip by 57pt

\begin{abstract}
The magnetic excitation spectrum of the $t$--$t'$--$J$-model is
studied in mean-field theory and compared to inelastic
neutron-scattering (INS) experiments on YBa$_2$Cu$_3$O$_{6+y}$ (YBCO)
and Bi$_2$Sr$_2$CaCu$_2$O$_{8+\delta}$ (BSCCO) superconductors. Within
the slave-particle formulation the dynamical spin response is
calculated from a renormalized Fermi liquid with an effective
interaction $\sim J$ in the magnetic particle--hole channel. We obtain
the so-called ``41\,meV resonance'' at wave vector $(\pi,\pi)$ as a
collective spin-1 excitation in the d-wave superconducting state. It
appears sharp (undamped), if the underlying Fermi surface is hole-like
with a sufficient next-nearest-neighbor hopping $t'<0$\,. The
double-layer structure of YBCO or BSCCO is not important for the
resonance to form. The resonance energy $\omega_{res}$ and spectral
weight at optimal doping come out comparable to experiment. The
observed qualitative behavior of $\omega_{res}$ with hole filling is
reproduced in the underdoped as well as overdoped regime. A second,
much broader peak becomes visible in the magnetic excitation spectrum
if the 2D wave-vector is integrated over. It is caused by excitations
across the maximum gap, and in contrast to the resonance its energy is
almost independent of doping. At energies above or below
$\omega_{res}$ the commensurate resonance splits into incommensurate
peaks, located off $(\pi,\pi)$\,. Below $\omega_{res}$ the intensity
pattern is of ``parallel'' type and the dispersion relation of
incommensurate peaks has a negative curvature. This is in accordance
with recent INS experiments on YBCO\,.

\end{abstract}
\pacs{PACS numbers: 71.10.Fd, 74.25.Ha, 74.72.Bk, 75.20.Hr}

]
\narrowtext

\newlength{\mywidth}
\renewcommand{\textfraction}{0.0}
\renewcommand{\topfraction}{1.0}
\renewcommand{\bottomfraction}{1.0}
\section{Introduction} \label{sec-intro}
The study of magnetic excitations plays an important r{\^o}le in the
ongoing attempt to understand the physics of High-Temperature
Superconductors. A key observation is the so-called ``41\,meV
resonance'' from inelastic neutron scattering (INS) experiments
\cite{ros91,moo93,fon95,bou96}\,. In superconducting optimally doped
YBa$_2$Cu$_3$O$_{6+y}$ (YBCO$_{6+y}$) a sharp peak occurs in the
magnetic structure factor at the antiferromagnetic (AF) wave vector
${\bf q}=(\pi,\pi)$ and energy $41$\,meV\,. It appears resolution
limited in energy and therefore is described as an (undamped)
$\delta$-peak. This is not expected for a d-wave superconductor, since
the density of states is finite and the resonance energy
$\omega_{res}\approx 40$\,meV is not small compared to $2\Delta^0$\,,
with the maximum gap $\Delta^0\sim 30\ldots 40$\,meV
\cite{rem-gap}\,. When temperature is raised through $T_c\approx 93$\,K
into the normal state, the resonance vanishes. The main effect of
underdoping \cite{dai96,fon97,fong00,daimook01} on the resonance in
the superconducting state is a continuous reduction of its
energy, as far as $\omega_{res}\approx 24$\,meV for the most
underdoped samples $T_c\approx 50$\,K\,. The resonance also gains
spectral weight with underdoping. In contrast to the optimally doped
case it persists into the normal state above $T_c$\,, where the
pseudo-gap regime is found. Recently a resonance has also been
observed \cite{fong99b,he01,mesot01pre} in 
Bi$_2$Sr$_2$CaCu$_2$O$_{8+\delta}$ (BSCCO$_{8+\delta}$)\,. Its energy
43\,meV in the optimally doped sample is comparable to the case of
YBCO\,. If experiments on YBCO and BSCCO are put together,
$\omega_{res}$ seems to follow $T_c$\,, i.e., it is maximal for
optimal doping and is reduced in under- as well as overdoped
compounds \cite{daimook01,he01}\,.

In this paper we report theoretical calculations of the magnetic
excitation spectrum in YBCO and BSCCO\,. Our starting point is the
doped Mott insulator, described by the $t$--$J$-model. We follow the
standard procedure of introducing auxiliary ``slave'' particles and
treating the problem in mean-field theory. In the resulting effective
theory the d-wave superconducting phase and the pseudo-gap regime of
underdoped systems are represented as spin--charge separated
states. The dynamical magnetic susceptibility is obtained from a Fermi
liquid of strongly renormalized quasi particles that carry the
spin. The ``41\,meV resonance'' is interpreted as a collective spin-1
excitation, it arises from a particle--hole (ph) bound state of these
quasi particles. We find results in good qualitative agreement with
the neutron-scattering experiments. In particular, the behavior of the
resonance energy $\omega_{res}$ with hole filling is reproduced for
underdoped as well as overdoped systems, and we obtain reasonable
absolute values for $\omega_{res}$ and spectral weight of the
resonance at optimal doping. Our findings are discussed in detail from
the doping dependent band structure of the quasi particles.

The concept of the resonance coming from a ph-bound state has been put
forward in several studies using ``slave''-particle schemes for
$t$--$J$ and Hubbard models
\cite{tan91,tan94,zhalev93,liu95,ste94}\,, a Hubbard-operator
technique \cite{onu95}\,, approaches based on BCS theory
\cite{mak94,maz95,bul96,salsch98,norm01}\,, and self-consistent
treatments of spin fluctuations in the Hubbard model (FLEX)
\cite{paobic95,dahm98,takmoriya98} or spin-fermion model
\cite{abachu99,morr98}\,. In the SO(5) approach
\cite{dem95,zha97}\,, on the other hand, the resonance is a result of a
bound state in the spin-triplet particle--particle (pp) channel, which
couples to the magnetic ph-channel in the superconducting state. In
Ref.\ \onlinecite{bri98} we studied the contribution from the pp-channel
within the present slave-particle scheme and concluded that it cannot
give rise to a resonance below $2\Delta^0$ unless unreasonable
parameters are used. A similar conclusion has been given in Ref.\
\onlinecite{tchnorchu01}\,. 

The resonance is connected to incommensurate structures in wave-vector
space. Above and below the resonance energy $\omega_{res}$ a splitting
of the single peak at ${\bf q}=(\pi,\pi)$ into four peaks slightly
displaced from $(\pi,\pi)$ is observed
\cite{stern94,bou97,mook98,arai99,bou00,mook00}\,.
When energy is raised from $\omega_{res}$ these follow a dispersion
$\omega({\bf q})$ similar to AF spin waves; however, the peaks are
very broad. Below $\omega_{res}$ four well separated peaks are
visible, which move away from $(\pi,\pi)$ with decreasing energy and
hence are described by an `upside-down' dispersion. Interestingly
these peaks are displaced from the AF wave vector in direction of the
$(\pi,0)$ or $(0,\pi)$ points, rotated from the nodal directions by
$45^\circ$\,. This is the same ``parallel'' type of incommensurability
as is known from the La$_{2-x}$Sr$_x$CuO$_{4+y}$ (LSCO) family of
compounds \cite{mas93,yam95}\,, where it has been brought into
connection to the so-called ``stripes''. In the present work we do
not consider the possibility of a combined ordering of spin and charge
into quasi one-dimensional (stripe-like) structures. Nevertheless,
below the resonance energy we obtain an incommensurate pattern of
parallel type. This is due to a ``dynamic nesting'' mechanism in the
superconducting state that enhances the intensity at these particular
points in wave-vector space. The dispersion relations of
incommensurate peaks are traced back to two particle--hole excitation
thresholds that vary differently with wave-vector.

Recently the magnetic response has also been studied by averaging the
neutron-scattering data over the 2D Brillouin zone
\cite{fong00,bou97,daimook99}\,. Besides the resonance, the resulting
local magnetic excitation spectrum ${\rm Im}\chi_{2D}(\omega)$ shows a
second, broad feature at an energy above the resonance, which depends
only weakly on the doping level. Within our calculation this feature
is naturally explained from particle--hole excitations across the
maximum d-wave gap $\Delta^0$\,. Their energy $\omega_{hump}\lesssim
2\Delta^0$ comes out almost independent of doping.

The paper is organized as follows: In Sections \ref{sec-mean} to
\ref{sec-effint} the mean-field theory for the $t$--$t'$--$J$-model is
derived and some basic implications are reviewed. The magnetic
resonance at the AF wave vector $(\pi,\pi)$ is considered in Section
\ref{sec-reson} for a single CuO$_2$ plane. Section
\ref{sec-incomm} presents results for the magnetic response in wave-vector
space. We consider the crossover from commensurate to incommensurate
response and the dispersion of incommensurate neutron peaks. In
Section \ref{sec-bilay} we take into account that YBCO and BSCCO are
actually bi-layer materials with two coupled CuO$_2$ planes per unit cell.
The splitting of the susceptibility into two modes is calculated. The
single-layer model considered in the previous sections serves as an
effective model for the odd (``acoustic'') mode, where the resonance
is observed. In this section we also discuss the above-mentioned local
susceptibility ${\rm Im}\chi_{2D}(\omega)$\,. A summary is given in
Section \ref{sec-concl}\,.

Some of the results have been presented briefly in Refs.\
\cite{bri98,bri99}\,. Work of other authors is further referenced in
the respective sections.

\section{Model and mean-field theory} \label{sec-mean}
We study the $t$--$J$-model on a simple square lattice of Cu-3d orbitals
for each of the two CuO$_2$ layers in YBCO or BSCCO\,:
\begin{equation} \label{eqn-model}
  H = - \sum_{\nu,\nu',\sigma}t_{\nu\nu'}
          \widetilde{ c}^\dagger_{\nu \sigma}
          \widetilde{ c}_{\nu' \sigma} +
    \frac{1}{2}\sum_{\nu,\nu'}
    J_{\nu\nu'}\vec{S}_\nu \vec{S}_{\nu'}
    \;.
\end{equation}
In the subspace with no doubly occupied orbitals, the electron
operator on a Cu-lattice site $\nu$ is denoted 
$\widetilde{ c}_{\nu \sigma}$
with spin index $\sigma=\pm 1$\,; 
$\vec{S}_\nu$ is the spin-density operator. 
A Cu-site is specified through $\nu\equiv [i,l]$\,, where $i= 1\ldots
N_L$ indicates the Cu-position within one CuO$_2$-plane and $l=1,2$
selects the layer in the bi-layer sandwich. $t_{\nu\nu'}$ denotes the
effective intra- and inter-layer Cu--Cu-hopping matrix elements, and
$J_{\nu\nu'}$ the antiferromagnetic super exchange .

To deal with the constraint of no double occupancy, the standard
auxiliary-particle formulation is used, 
\begin{equation}  \label{eqn-decom}
  \widetilde{ c}_{\nu \sigma} = 
  b^\dagger_\nu f_{\nu \sigma}\,.
\end{equation}
The fermion $f^\dagger_{\nu \sigma}$ creates a singly occupied site
(with spin $\sigma$), the ``slave'' boson $b^\dagger_\nu$ an empty
one out of the (unphysical) vacuum
$b_\nu|0\rangle = f_{\nu \sigma}|0\rangle = 0$\,.
The constraint now takes the form
\begin{equation} \label{eqn-const}
  Q_\nu = 
    b^\dagger_\nu b_\nu + \sum_\sigma 
    f^\dagger_{\nu \sigma} f_{\nu \sigma} = 1
    \;.
\end{equation}
Using (\ref{eqn-const}), local operators can be written in
fermions, in particular the particle and spin density read
\begin{equation}  \label{eqn-inferm}
  n_\nu = 
    \psi^\dagger_{\nu} \psi_{\nu}
  \;\;,\;\;
  \vec{S}_\nu = 
    \frac{1}{2}
    \psi^\dagger_{\nu}\vec{\tau}\psi_{\nu}
\end{equation}
Here spinors
$\psi_\nu = \left({f_{\nu\uparrow}}\atop{f_{\nu\downarrow}}\right)$\,, 
$\psi^\dagger_\nu = 
   \left({f^\dagger_{\nu\uparrow}}\atop{f^\dagger_{\nu\downarrow}}\right)$
have been introduced, with Pauli matrices $\vec{\tau}$ and
$\hbar\equiv 1$\,. 

In order to derive a mean-field theory the constraint
(\ref{eqn-const}) is relaxed to its thermal average 
$\langle Q_\nu \rangle = 1$\,.
Together with the number $x$ of doped holes per Cu-site,
it fixes the fermion and boson densities to 
\begin{equation} \label{eqn-dens}
    (1 - x) = 
        \langle \psi^\dagger_{\nu}
        \psi_{\nu} \rangle 
    \;,\;\;\;\;
    x = 
      \langle b^\dagger_\nu b_\nu \rangle
    \;.
\end{equation}
These are adjusted by chemical potentials $\mu^b, \mu^f$\,. Using a
coherent-state path integral the partition function is now represented
by the action
\begin{equation}  \label{eqn-action}
  S = S^0 + S^t + S^J + S^h
\end{equation}
with
\begin{eqnarray*}
  S^0 & = &  
    \int_0^\beta\!\!{\rm d}\tau 
      \sum_\nu \left\{
      \bar{b}_\nu (\partial_\tau - \mu^b) b_\nu + 
      \bar{\psi}_{\nu} (\partial_\tau - \mu^f) \psi_{\nu}
      \right\}
    \\
  S^t & = &  
    - \int_0^\beta\!\!{\rm d}\tau
    \sum_{\nu,\nu'} t_{\nu\nu'}
    \bar{\psi}_{\nu} \psi_{\nu'} \bar{b}_{\nu'} b_\nu 
    \\
  S^J & = &
    \int_0^\beta\!\!{\rm d}\tau
    \frac{1}{2}\sum_{\nu,\nu'}
    J_{\nu \nu'} \vec{S}_\nu \vec{S}_{\nu'}
    \;\;\;,\;\;\;
  S^h = 
    - \int_0^\beta\!\!{\rm d}\tau
    \sum_\nu
    \vec{h}_\nu \vec{S}_\nu
\end{eqnarray*}
A magnetic source-field $\vec{h}\equiv \vec{h}_\nu(\tau)$ has been
added here, $\beta \equiv 1\,/\,k_BT$\,.

A mean-field decomposition of the interaction terms $S^t, S^J$
is achieved via Feynman's variational principle \cite{feynbook} for
the free energy $F$\,,
\begin{equation}  \label{eqn-feyn}
  \beta F \le \Psi[\widetilde{ S}]
    \;\;,\;\;
  \Psi[\widetilde{ S}] = 
  \langle S - \widetilde{ S} \rangle - \ln \widetilde{ Z}
\end{equation}
The effective action $\widetilde{ S}$ determines
$\widetilde{ Z} = 
    \int{\sc D}[\psi,\bar{\psi},b,\bar{b}]
    \exp(-\widetilde{ S})$
and thermal averages
\begin{displaymath}
 \langle \widehat{O} \rangle = 
    \frac{1}{\widetilde{ Z}} 
    \int{\sc D}[\psi,\bar{\psi},b,\bar{b}]
    \exp(-\widetilde{ S})\,\widehat{O}
\end{displaymath}
For $\widetilde{ S}$ we make the quadratic ansatz
\begin{eqnarray}  \label{eqn-effact}
  \widetilde{ S} & = & S^0 
    + \int{\rm d}1 {\rm d}2 \bigg\{
    \bar{\psi}_1 {\bf T}^f_{1 2} \psi_2
    \\
  & & \mbox{}  \nonumber
    + \bar{b}_{1} T^b_{1 2} b_{2}
    + \left( \bar{\psi}_1 {\bf A}_{1 2} \bar{\psi}_2 + h.c.\right)
    \bigg\}
    \\
  & & \mbox{}  \nonumber
    - \int{\rm d}1\, \vec{m}_1 \vec{S}_1
\end{eqnarray}
with a shorthand notation $1\equiv (\nu_1, \tau_1)$\,, $\int{\rm d}1
\equiv \sum_{\nu_1}\int_0^\beta\!\!{\rm d}\tau_1$ collecting site
$\nu$ and time $\tau$ indices. ${\bf T}^f_{12}$ and ${\bf A}_{12}$ are
matrices in spin space, e.g., ${\bf A}_{12}\equiv A_{12}^{\sigma_1
\sigma_2}\equiv A_{\nu_1 \nu_2}^{\sigma_1
\sigma_2}(\tau_1,\tau_2)$\,. $\widetilde{ S}$ consists 
of quadratic terms for fermions and bosons, which represent all
possible mean-field decouplings of the interactions $S^t$ and $S^J$\,.

The expectation value $\langle S - \widetilde{
S}\rangle$ in Eq.\ (\ref{eqn-feyn}) is calculated using Wick's
theorem, and from the vanishing variation $\delta\Psi[\widetilde{
S}]=0$ we obtain equations for the self-consistent parameters,
\begin{mathletters}  \label{eqn-mf}
\begin{eqnarray} 
  T^b_{1 2} & = &  \label{eqn-mf-tb}
    - t_{1 2} \langle \bar{\psi}_2 \psi_1 \rangle
    \\
  T_{1 2 }^{f\,\sigma \sigma'} & = &  \label{eqn-mf-tf}
    - t_{1 2} \langle \bar{b}_2 b_1 \rangle \delta_{\sigma \sigma'}
    \\
  & &  \nonumber\mbox{}
    - J_{1 2}\frac{1}{4} \sum_{\mu=1}^3
    \langle (\tau_\mu^T\bar{\psi}_2)^{\sigma'}\,
            (\tau_\mu\psi_1)^\sigma \rangle
    \\
  A_{1 2}^{\sigma \sigma'} & = &  \label{eqn-mf-a}
    - J_{1 2} \frac{1}{8} \sum_{\mu=1}^3
    \langle (\tau_\mu\psi_1)^\sigma\,
            (\tau_\mu\psi_2)^{\sigma'} \rangle
    \\
  \vec{m}_1 & = &  \label{eqn-mf-m}
    \vec{h}_1 - \int{\rm d}2\, J_{1 2}\langle\vec{S}_2\rangle
\end{eqnarray}
\end{mathletters}
$\tau_\mu$ denotes a Pauli matrix, $\tau_\mu^T$ its transpose. The
effective hopping $T^b$ of bosons as well as the first contribution to
the hopping $T^f$ of fermions stem from the decoupling of $\langle S^t
\rangle$ in Eq.\ (\ref{eqn-feyn})\,. The Heisenberg term $\langle S^J
\rangle$ is factorized through Wick's theorem into contributions to
the local magnetic field $\vec{m}$\,, the fermion's hopping $T^f$ and
pairing amplitude $A$\,. These correspond to analyzing $S^J$ in the
direct particle--hole (ph) channel of fermions, the exchange ph, and
the particle--particle channel,
respectively\cite{rem-hstrafo}\,. Further below, when the source field
$\vec{h}$ is set to zero, these will be restricted to a Resonating
Valence-Bond (RVB) amplitude
$T^f_{\nu \nu'} \sim
   \langle \bar{f}_{\nu \uparrow} f_{\nu' \uparrow} \rangle$
and spin-singlet pairing
$A_{\nu \nu'} \sim
   \langle {f}_{\nu \uparrow} {f}_{\nu' \downarrow} 
     \rangle$\,.

\begin{figure}[h]
  \mywidth= 0.2\hsize
  \begin{displaymath}
    \begin{array}{rcl} \displaystyle 
        \chi^{irr} & = &  \displaystyle 
        \parbox{1.2\mywidth}{\includegraphics[width=1.2\mywidth]{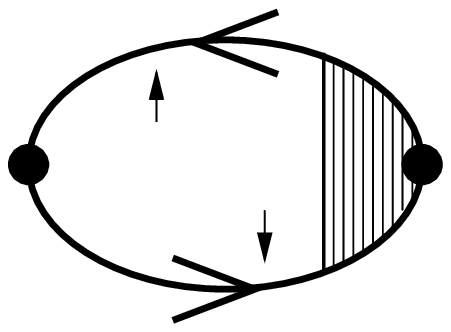}}
        \; + \;
        \parbox{1.2\mywidth}{\includegraphics[width=1.2\mywidth]{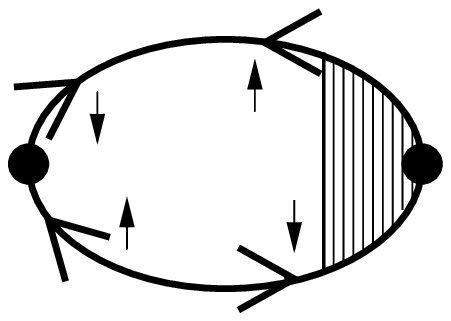}}
        \\ \\ \displaystyle 
        \parbox{0.68\mywidth}{\includegraphics[width=0.68\mywidth]{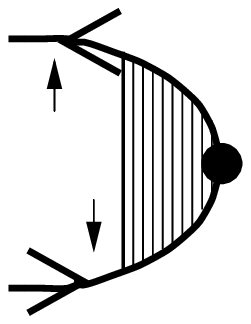}}
        & = &
        \parbox{0.68\mywidth}{\includegraphics[width=0.68\mywidth]{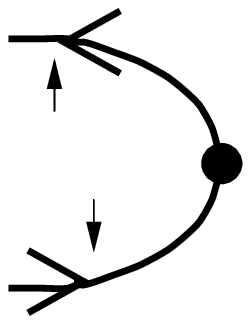}}
        \; + \;
        \parbox{\mywidth}{\includegraphics[width=\mywidth]{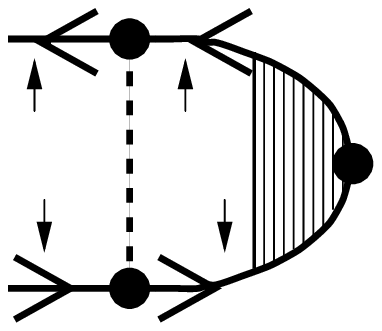}}
        \; + \;
        \parbox{1.15\mywidth}{\includegraphics[width=1.15\mywidth]{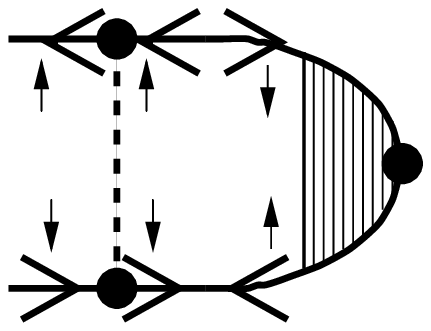}}
    \end{array}
  \end{displaymath}
  \caption[\ ]{
    {\bf Top:} Irreducible part $\chi^{irr}$ for vanishing source
  field $\vec{h}=0$\,. Full lines are fermion Green's functions, boson
  excitations do not enter $\chi^{irr}$ at mean-field level.
    {\bf Bottom:} Bethe--Salpeter equation for the vertex function in
  $\chi^{irr}$\,, dashed lines stand for the Heisenberg interaction
  $J$\,. The bare vertex (i.e., the unrenormalized bubble) 
  represents $\langle \vec{S}_1 \vec{S}_2\rangle^{conn}$ in Eq.\
  (\ref{eqn-irr})\,.
    }  
  \label{fig-chiirr}
\end{figure}

The approximate free energy is the functional $\Psi$ at
its stationary point, 
$\beta F^{appr} = \Psi[\widetilde{ S}^{stat}]$\,,
with the action $\widetilde{ S}^{stat}$ given by Eq.\
(\ref{eqn-effact}) and Eqs.\ (\ref{eqn-mf})\,.  The dynamical
magnetic susceptibility then follows with
$\delta\Psi[\widetilde{ S}^{stat}]\,/\,\delta\vec{h}_1 = 
  - \langle \vec{S}_1 \rangle$
as
\begin{displaymath}
  \tensor{\chi}_{12} = 
    - \frac{\delta^2 \Psi[\widetilde{ S}^{stat}]}
           {\delta\vec{h}_1 \delta\vec{h}_2}
    = \frac{\delta}{\delta\vec{h}_1} \langle\vec{S}_2\rangle
    = \int{\rm d}3\,
      \frac{\delta\langle\vec{S}_2 \rangle}{\delta\vec{m}_3}
      \frac{\delta\vec{m}_3}{\delta\vec{h}_1}
\end{displaymath}
Using Eq.\ (\ref{eqn-mf-m}) we obtain the usual `RPA-like'
expression; in a matrix notation it reads
\begin{equation}  \label{eqn-rpa}
  \chi = \left[ 1 + J \chi^{irr} \right]^{-1}
    \chi^{irr}
\end{equation}
The irreducible part is identified as
\begin{equation}  \label{eqn-irr}
  \tensor{\chi}^{irr}_{12} = 
    \frac{\delta\langle\vec{S}_2\rangle}{\delta\vec{m}_1}
    = \langle \vec{S}_1 \vec{S}_2 \rangle^{conn} +
      \left( \frac{\delta\langle\vec{S}_2\rangle}{\delta\vec{m}_1} 
        \right)_{impl}
\end{equation}
The 1st term on the r.h.s.\ comes from the $\vec{m}$ which appears
explicitly in $\langle\vec{S}_2\rangle$ through $\widetilde{ S}$\,,
Eq.\ (\ref{eqn-effact})\,. For a vanishing source field $\vec{h}=0$ it
is given by the unrenormalized fermion bubble contained in Fig.\
\ref{fig-chiirr}\,. Since the operator $\vec{S}$ involves only
fermions, no boson excitation occur in $\chi^{irr}$\,.  The 2nd term
in Eq.\ (\ref{eqn-irr}) stands for all contributions from the implicit
$\vec{m}$-dependence of $\langle\vec{S}_2\rangle$ through the other
mean-field parameters $T^b, T^f, A$\,. Using Eqs.\
(\ref{eqn-mf-tb}--\ref{eqn-mf-a})\,, it gives rise to vertex
corrections. These are shown in Fig.\ \ref{fig-chiirr} (bottom).

In the following we set $\vec{h}=0$ and consider paramagnetic phases
$\vec{m}=0$\,, which are symmetric with respect to lattice
translations within a CuO$_2$-layer and exchange of the layers. In
going to wave-vector space, the site index $\nu\equiv [i,l]$\,, with
in-plane site $i=1\ldots N_L$ and layer index $l=1,2$ is replaced by
the wave-vector $p\equiv ({\bf k}, p_z)$\,. That is,
\begin{displaymath}
  f_{il \sigma} = 
    \frac{1}{\sqrt{2 N_L}} \sum_k e^{i(k_x i_x + k_y i_y)}
    \sum_{p_z= 0, \pi}e^{i p_z l} f_{p \sigma} 
\end{displaymath}
and similar for boson operators. Here ${\bf k}$ runs over the usual 2D
Brillouin zone, and $p_z= 0, \pi$ corresponds to even, odd linear
combination of layer orbitals. 

The exchange interaction is decomposed as
\begin{equation}  \label{eqn-jj}
  J_{\nu\nu'} \equiv
    J^{ll'}_{ij} = 
    \delta_{ll'} J \delta_{<i,j>} + 
    (1 - \delta_{ll'}) J^\perp \delta_{ij}
\end{equation}
It consists of an intra-layer component $J$ for nearest neighbors
\mbox{$<i,j>$} and an inter-layer coupling $J^\perp$ for $i=j$\,.
Thus the pairing $A$ of fermions, Eq.\ (\ref{eqn-mf-a}), involves an
intra-layer part, which we restrict to singlet pairing with d-wave
symmetry and equal amplitude and phase in both layers:
$\langle f_{l\,i \uparrow} f_{l\,i + \hat{x} \downarrow} \rangle = 
   - \langle f_{l\,i \uparrow} f_{l\,i + \hat{y} \downarrow} \rangle$\,,
$l=1,2$\,. $A$ contains as well an inter-layer amplitude
$\langle f_{1 i \uparrow} f_{2 i \downarrow} \rangle$\,. 
In $p$-space Eq.\ (\ref{eqn-mf-a}) then becomes
$A_{12}^{\sigma \sigma'} \to 
  A^{\sigma \sigma'}(p) =
  \sigma \delta_{-\sigma'\,\sigma} \Delta_p$\,,
with the gap function
\begin{equation}  \label{eqn-gapfun}
  \Delta_p = 
    \frac{\Delta^0}{2}[\cos(k_x) - \cos(k_y)] + 
    \Delta^{\perp 0} e^{i p_z}
\end{equation}
and the maximum in-plane and inter-plane gap \\
$\Delta^0= 
   \frac{3}{2}J\widehat{\Delta}$\,, 
$\Delta^{\perp 0}= 
   \frac{3}{8}J^\perp\widehat{\Delta}^\perp$\,,
\begin{mathletters}  \label{eqn-pair}
\begin{eqnarray}
  \widehat{\Delta} & = &  \label{eqn-pair-par}
    \langle f_{1\,i\uparrow} f_{1\,i+\hat{x}\downarrow} \rangle
    - \langle f_{1\,i\downarrow} f_{1\,i+\hat{x}\uparrow} \rangle
    \\
  \widehat{\Delta}^\perp & = &  \label{eqn-pair-perp}
    \langle f_{1\,i\uparrow} f_{2\,i\downarrow} \rangle
    - \langle f_{1\,i\downarrow} f_{2\,i\uparrow} \rangle
\end{eqnarray}
\end{mathletters}

As will be explained at the end of this section, in the interesting
range of temperature and hole filling $x$ the bosons may be treated as
almost condensed. That is, the hopping rate
$\langle \bar{b}_\nu b_{\nu'} \rangle \approx
   \langle \bar{b}_\nu b_\nu \rangle = x$
is independent of $\nu, \nu'$ and given by the hole density $x$ via
Eq.\ (\ref{eqn-dens})\,.  The first term in the fermion hopping Eq.\
(\ref{eqn-mf-tf}) becomes
$- t_{\nu \nu'}\langle \bar{b}_\nu b_{\nu'} \rangle \to
  - x t_{\nu \nu'}$\,.
It describes the propagation of fermions with the small probability
$x$ of finding an empty site. The second term in Eq.\
(\ref{eqn-mf-tf}) involves induced hopping amplitudes on
nearest-neighbor bonds, which we assume equal in amplitude and phase
on each bond within a layer (uniform RVB),
$\langle \bar{f}_{l\,i\uparrow} f_{l\,i+\hat{x}\uparrow} \rangle
   = \langle \bar{f}_{l\,i\uparrow} f_{l\,i+\hat{y}\uparrow} \rangle$\,,
$l= 1,2$\,. 
The fermion hopping Eq.\ (\ref{eqn-mf-tf}) now turns into
$T^{f\,\sigma \sigma'}_{12} \to T^{f\,\sigma \sigma'}(p) =
  \delta_{\sigma \sigma'} T^f_p$\,,
\begin{eqnarray}
  T^f_p & = &  \label{eqn-fhop}
    - 2 \widetilde{ t}[\cos(k_x) + \cos(k_y)]
    \\ 
  & &  \nonumber
    \mbox{} - 4 \widetilde{ t}'\cos(k_x) \cos(k_y)
    - \widetilde{ t}^\perp({\bf k}) e^{i p_z}
\end{eqnarray}
with
\begin{eqnarray}  
  \widetilde{ t} & = &  \label{eqn-modt}
    x\,t + \frac{3}{8} J \widehat{\chi}
    \;,\;\;\;
  \widetilde{ t}'= x\,t'
    \\
  \widetilde{ t}^\perp({\bf k}) & = &  \nonumber
    x\,t^\perp({\bf k}) + \frac{3}{8}J^\perp\widehat{\chi}^\perp
\end{eqnarray}
and
\begin{mathletters}  \label{eqn-hopamp}
\begin{eqnarray}
  \widehat{\chi} & = &  \label{eqn-hopamp-par}
    \langle \bar{f}_{1\,i\uparrow} f_{1\,i+\hat{x}\uparrow} \rangle
    + \langle \bar{f}_{1\,i\downarrow} f_{1\,i+\hat{x}\downarrow} \rangle
    \\
  \widehat{\chi}^\perp & = &  \label{eqn-hopamp-perp}
    \langle \bar{f}_{1\,i\uparrow} f_{2\,i\uparrow} \rangle
    + \langle \bar{f}_{1\,i\downarrow} f_{2\,i\downarrow} \rangle
\end{eqnarray}
\end{mathletters}
For the bare hopping elements we assumed a nearest and next-nearest
neighbor overlap $t$ and $t'$ within a layer, and an inter-layer
\cite{chasud93,okand94} hopping 
$t^\perp({\bf k}) = 
  2 t^\perp [\cos(k_x) - \cos(k_y)]^2 + t^\perp_0$\,.
For bosons the effective hopping is derived similarly,
$T^b_{12}\to T^b_p$\,.
The result is given at the end of this section.

The mean-field Hamiltonian Eq.\ (\ref{eqn-effact}) now reads
\begin{eqnarray}  \label{eqn-mfact}
  \widetilde{ S} & = &
    \int_0^\beta\!\!{\rm d}\tau \sum_p\bigg\{
      \bar{b}_p(\partial_\tau + \Omega_p) b_p
      + \sum_\sigma 
      \bar{f}_{p \sigma}(\partial_\tau + \varepsilon_p) f_{p \sigma}
    \\
  & &  \nonumber
    \mbox{} + \left[
    \Delta_p(\bar{f}_{p\uparrow}\bar{f}_{-p\downarrow} 
         - \bar{f}_{p\downarrow}\bar{f}_{-p\uparrow}) + h.c.\ 
    \right] \bigg\}
\end{eqnarray}
It consists of free bosons with dispersion 
$\Omega_p = T^b_p - \mu^b$
and BCS-fermions with 
$\varepsilon_p = T^f_p - \mu^f$ 
and gap function Eq.\ (\ref{eqn-gapfun})\,.  After Bogoliubov
transformation we obtain `quasi-fermion' energies 
$E_p= \sqrt{\varepsilon_p^2 + \Delta_p^2}$\,, 
and the mean-field equations (\ref{eqn-pair}), (\ref{eqn-hopamp})
become 
\begin{mathletters}  \label{eqn-mfeq}
\begin{eqnarray}
  \left( \widehat{\chi} \atop \widehat{\chi}^\perp \right)
    & = &  \label{eqn-mfeq-chi}
    - \frac{1}{2N_L}\sum_p
    \left( \gamma({\bf k})/2 \atop e^{i p_z} \right)
    \frac{\varepsilon_p}{E_p} \tanh(\beta E_p / 2)
    \\
  \left( \widehat{\Delta} \atop \widehat{\Delta}^\perp \right)
    & = &  \label{eqn-mfeq-del}
    \frac{1}{2N_L}\sum_p
    \left( \varphi({\bf k})/2 \atop e^{i p_z} \right)
    \frac{\Delta_p}{E_p} \tanh(\beta E_p / 2)
    \\
  x & = &  \label{eqn-mfeq-fill}
    \frac{1}{2N_L}\sum_p \frac{\varepsilon_p}{E_p}\tanh(\beta E_p/2)
\end{eqnarray}
\end{mathletters}
with phase factors
$\gamma({\bf k}) = \cos(k_x) + \cos(k_y)$\,,
$\varphi({\bf k}) = \cos(k_x) - \cos(k_y)$\,.
The last equation is the particle number constraint
(\ref{eqn-dens})\,. 

The magnetic susceptibility Eq.\ (\ref{eqn-rpa}) is isotropic in
spin-space for $\vec{h}=\vec{m}=0$\,, and takes the usual form 
\begin{equation}  \label{eqn-sus}
  \chi_p(\omega) = 
    \frac{\chi^{irr}_p(\omega)}{1 + J_p \chi^{irr}_p(\omega)}
\end{equation}
where $J_p$ is obtained from Eq.\ (\ref{eqn-jj}) as
\begin{equation}  \label{eqn-jjp}
  J_p = 
    2J\left[ \cos(q_x) + \cos(q_y) \right] + 
    e^{i p_z}J^\perp
\end{equation}
In experiment the magnetic response $\chi^{meas}$ is measured 
as a function of the wave vector $({\bf q}, q_z)$\,,
which spans the 3D Brillouin zone of the bi-layer material. It is
given by \cite{rem-prlchi}
\begin{eqnarray}
  \lefteqn{\chi^{meas}({\bf q}, q_z, \omega) = }  \label{eqn-chimeas}
    \\
  & &  \nonumber
    \;\;\; (g\mu_B)^2 \left[
    \left.\chi_{p}(\omega)\right|_{p_z=0} \cos^2(\frac{d}{2}q_z) +
    \left.\chi_{p}(\omega)\right|_{p_z=\pi} \sin^2(\frac{d}{2}q_z)
    \right]
\end{eqnarray}
with $p=({\bf q}, p_z)$\,. $d$ denotes the spacing of layers in the
double-layer.  The even ($p_z=0$) and odd ($p_z=\pi$) mode
susceptibilities
correspond to the in-phase and anti-phase combination of spin
fluctuations in the planes. For the irreducible part Eq.\
(\ref{eqn-irr}) we take the bare bubble 
$\chi^{irr}_{\nu\nu'}(\tau,\tau') = 
  \langle S^z_\nu(\tau) S^z_{\nu'}(\tau') \rangle^{conn}$\,.
The vertex corrections depicted in the bottom of Fig.\
\ref{fig-chiirr} can be savely ignored. As we have discussed in Ref.\
\onlinecite{bri98} they have no significant effect in the interesting energy
range $0\le\omega\le 2\Delta^0$\,. With the effective Hamiltonian Eq.\
(\ref{eqn-mfact}) we get the expression known from BCS theory
\begin{eqnarray}
  \lefteqn{ \chi^{irr}_p(\omega) = 
    \frac{1}{2N_L}\sum_{\tilde{p}}\sum_{s, s'=\pm 1} }   \label{eqn-bubb}
    \\
  & &   \nonumber
    \;\;\;\;\;\;
    \frac{1}{8}\left[ 1 + 
      ss'\frac{\varepsilon \varepsilon' + \Delta\Delta'}{E E'}
               \right]
    \frac{f(s'E') - f(s E)}{\omega + sE - s'E' + i0_+}
\end{eqnarray}
Here 
$\varepsilon\equiv \varepsilon_{\tilde{p}}$\,,
$\varepsilon'\equiv \varepsilon_{\tilde{p}+p}$\,,
and similar for $\Delta$\,, $E$\,. $f$ denotes the Fermi
function. 

We close this section with a remark on Bose condensation. From Eq.\
(\ref{eqn-mf-tb}) the boson dispersion is
$\Omega_p = - 2 t \widehat{\chi} [\cos(k_x) + \cos(k_y)] - \mu^b$\,,
where for simplicity $t'=t^\perp=0$\,. Near the band minimum $k=0$
this becomes 
$\Omega_p \approx \overline{\Omega} + k^2\,/\,2m^b$\,,
with the mass $1\,/\,m^b = 2 t \widehat{\chi}$\,. From the solution of
Eqs.\ (\ref{eqn-mfeq}) we get values around $\widehat{\chi} \approx
0.4$\,, i.e., $m^b \approx 1/t$\,. In two dimensions free bosons do not
condense at finite temperature $T>0$\,, however, the correlation
length of the propagator 
$\langle \bar{b}_\nu b_{\nu'} \rangle$
grows exponentially for $T$ below 
$T^0_{BE} = 2\pi x\,/\,m^b \approx 2\pi x t$\,.
In the $x$ and $T$ range we are interested in, $T\ll
T^0_{BE}$\,, and the bosons can be considered almost condensed, i.e.,
$\overline{\Omega}\to 0$ and 
$\langle \bar{b}_\nu b_{\nu'} \rangle \approx
   \langle \bar{b}_\nu b_\nu \rangle = x$
for any $\nu, \nu'$\,. 

\section{Phase diagram} \label{sec-phase}
\begin{figure}[h]
  \centerline{
     \includegraphics[width=0.9\hsize]{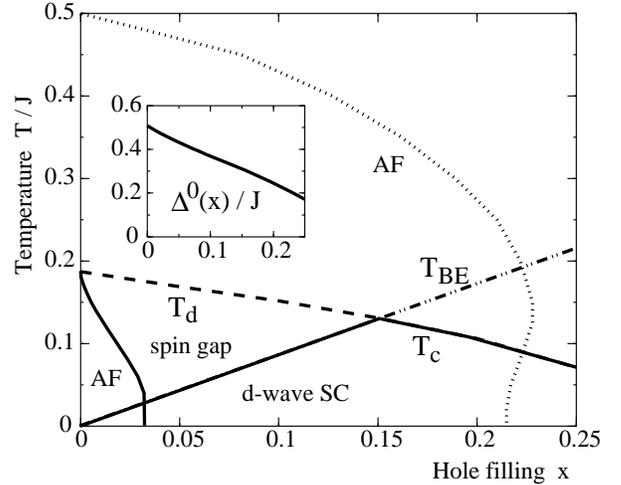} }
  \caption[\ ]{
    {\bf Main figure:} Mean-field phase diagram
    for a single CuO$_2$-layer with $t= 2J$\,. $T_d$ and
    $T_{BE}$ denote transition temperatures to d-wave pairing of
    fermions and condensation of bosons, respectively. $T_c= \min\{T_d,
    T_{BE}\}$ is the physical $T_c$ for bulk d-wave superconductivity. The
    lines labeled AF indicate the magnetic instability of the bare
    (dotted line) and renormalized theory (continuous), see text, Sec.\
    \ref{sec-effint}\,. 
    {\bf Inset:} Maximum gap $\Delta^0$ of fermions as function of
    hole filling at $T= 0$\,. 
    }
  \label{fig-phase}
\end{figure}

The slave-boson mean-field theory has been put forward in numerous
papers \cite{bza87,ruchir87,kot88,suzfuk88,affmar88}\,, originating in
the Resonating Valence-Bond (RVB) idea \cite{rvbphil}\,. In this
section we review the phase diagram and briefly discuss some
experimental implications in the superconducting phase at $T\to
0$\,. For simplicity a single CuO$_2$-layer is considered, with $t=
2J$\,. Fig.\ \ref{fig-phase} shows the phase diagram, derived from the
numerical solution of Eqs.\ (\ref{eqn-mfeq})\,. It resembles those
given in the literature \cite{fuk92,naglee92}\,, except at very small
doping $x\to 0$\,, where our assumption of almost condensed bosons
becomes incorrect. We also ignored the staggered-flux phase reported
\cite{ubblee92} for small $x$\,. These simplifications do not affect 
our results for the magnetic excitations. 

Ignoring the lines labeled `AF' for the moment, the phase diagram
shows two transition temperatures $T_d$ and $T_{BE}$\,. At
temperatures $T>T_d$ the fermions move in the CuO$_2$-plane with an
effective dispersion
$\varepsilon({\bf k})= - 2 \widetilde{ t}[\cos(k_x) +\cos(k_y)] -
    \mu^f$\,, 
where $\widehat{\chi}$ in
$\widetilde{ t}= x\,t + \frac{3}{8}J\widehat{\chi}$
is finite (uniform RVB phase). At $T= T_d$ they undergo a transition
to the d-wave paired state with order parameter
$\Delta({\bf k})= 
  \frac{\Delta^0}{2}[\cos(k_x) - \cos(k_y)]$\,.
This spin-gap phase is characterized by a gap $\Delta({\bf k})$ for
spin excitations, and is associated with the spin-gap (or pseudo-gap)
regime observed in the normal state of underdoped cuprates. The
bosons, on the other hand, show Bose condensation
$\langle b_\nu \rangle\ne 0$
at $T= T_{BE}$\,. Bulk superconductivity is present for $T< T_c=
\min\{T_d, T_{BE}\}$\,, where 
$\langle c_{\nu\uparrow} c_{\nu'\downarrow} \rangle = 
    \langle b^\dagger_\nu \rangle
    \langle b^\dagger_{\nu'} \rangle
    \langle f_{\nu\uparrow} f_{\nu'\downarrow} \rangle \ne 0$\,. 
In two dimensions $T_{BE}$ is identified with $T_{BE}^0= 2\pi x t$\,,
which yields a very large slope of the $T_{BE}$ line in the phase
diagram. Fluctuations of gauge fields around the mean-field solution
are expected to reduce $T_{BE}^0$ to reasonable values
\cite{naglee92}\,. The $T_{BE}$ line sketched in Fig.\ \ref{fig-phase}
corresponds to that situation, with a maximum $T_c$ at an optimal
doping value $x_{opt}\approx 0.15$\,. The similarity of spin-gap and
superconducting (SC) phase in mean-field theory naturally explains the
persistence of the magnetic resonance into the spin-gap regime,
although line-shape (damping) effects are missing.

In the following we focus on the SC state at $T\to 0$\,, which is
reasonably reproduced by mean-field theory: In cuprate superconductors
the underdoped region $x< x_{opt}$ shows unusual behavior of $T_c$\,,
the superfluid density, and the maximum gap as function of hole
filling. The superconducting $T_c$ increases with doping,
$T_c=T_{BE}\sim x$\,. The superfluid density $\rho^s= n^s/m$ is given
by the condensate density $\rho^b= n^b/m^b\sim x$ of bosons for small
$x$\,, according to the Ioffe--Larkin formula
\cite{iof89}\,. Thus $\rho^s\sim x$\,, and the well known experimental
observation \cite{uem89} $\rho^s\sim T_c$ follows naturally. Tunneling
and photoemission spectra are described by the Green's function
$G_{\nu \nu'}(\tau,\tau')= 
  \langle {\cal T}_\tau 
   \widetilde{ c}_{\nu \uparrow}(\tau) 
   \widetilde{ c}^\dagger _{\nu' \uparrow}(\tau') \rangle$\,, 
where $\widetilde{ c}$ is expressed by Eq.\ (\ref{eqn-decom})\,.  In
the superconducting state, where bosons are condensed in
${\bf k}= 0$\,, $G$ splits into a coherent and incoherent part,
$G= x\,{\cal G}^{ferm} + G^{incoh}$\,,
where ${\cal G}^{ferm}$ is the propagator of fermions.  Thus the
superconducting gap is given by the d-wave pairing gap of the
fermions. Its doping dependence at $T=0$ is shown in the inset of
Fig.\ \ref{fig-phase}\,. When $x$ is reduced from $x_{opt}$ the
maximum gap actually increases (whereas $T_c$ decreases), as is seen
in experiment \cite{ren98,miya98,loe96,din96}\,. At optimal doping
mean-field theory gives $\Delta^0\approx 0.3 J\approx 40\,$meV\,,
which compares reasonably with experimental values
\cite{rem-gap}\,. On the overdoped side $x> x_{opt}$\,, $T_c= T_d$\,,
and we get the BCS-like result $\Delta^0\sim T_c$\,.

Recently an alternative slave-boson formulation has been proposed
\cite{wen96,lee98}\,, which extends the SU(2) symmetry in
particle--hole space of the 1/2-filled model \cite{aff88} to the
hole-doped case. Within mean-field theory the superconducting state at
$T\to 0$ appears to be similar to the more conventional U(1)
formulation we are using here, in particular is the magnetic
spin-response the same. 

\section{Effective interaction} \label{sec-effint}
So far we have not considered the possibility of antiferromagnetic
(AF) order. It is known that wide areas of the mean-field phase
diagram are unstable to AF order
\cite{leefeng88,gros89,inui88,inaba96}\,. We determine the AF phase
boundary from the diverging correlation length $\xi_{AF}$\,, which is
extracted from the static ($\omega=0$) susceptibility. For a single
CuO$_2$-layer Eqs.\ (\ref{eqn-sus})--(\ref{eqn-bubb}) reduce to
\begin{equation}  \label{eqn-singlrpa}
  \chi({\bf q},\omega)= 
    \frac{\chi^{irr}({\bf q},\omega)}
         {1 + 2 J[\cos(q_x) + \cos(q_y)]\, \chi^{irr}({\bf q},\omega)}
\end{equation}
with $\chi$ in units \cite{rem-prlchi} of $(g\mu_B)^2$\,.  The
irreducible part
$\chi^{irr}({\bf q},\omega)= 
   \chi^{irr}_p(\omega)$
is given by Eq.\ (\ref{eqn-bubb}) with $p= ({\bf q}, p_z)$\,,
$\widetilde{p}= ({\bf k}, p_z)$ and arbitrary $p_z$\,. In Eq.\
(\ref{eqn-bubb}), the internal summation is now over the 2D Brillouin
zone,
$\frac{1}{2N_L}\sum_{\tilde{p}} \to \frac{1}{N_L}\sum_{\bf k} =
   \int_{-\pi}^{\pi}\!{\rm d}^2 k\,/\,(2\pi)^2$\,, 
and dispersion and gap function become
\begin{eqnarray}
  \varepsilon & \equiv &  \nonumber
    \varepsilon({\bf k}) = T^f_{\tilde{p}} - \mu^f
    \;\; \mbox{with} \;\; \widetilde{ t}^\perp=0
    \;;\;\;
    \varepsilon'\equiv \varepsilon({\bf k}+{\bf q})
    \\
  \Delta & \equiv &  \label{eqn-singldisp}
    \Delta({\bf k}) = \Delta_{\tilde{p}}
    \;\; \mbox{with} \;\; \Delta^{\perp 0}= 0
    \;;\;\;
    \Delta'\equiv \Delta({\bf k}+{\bf q})
    \\
  E & \equiv &  \nonumber
    E({\bf k})= \sqrt{\varepsilon({\bf k})^2 + \Delta({\bf k})^2}
    \;;\;\;
    E'\equiv E({\bf k}+{\bf q})
\end{eqnarray}
$\Delta_p$\,, $T^f_p$ have been given in Eqs.\
(\ref{eqn-gapfun},\ref{eqn-fhop},\ref{eqn-modt})\,.  At the N{\'e}el
wave-vector ${\bf Q}=(\pi,\pi)$ the static susceptibility takes the
value $\chi_{AF}= \chi({\bf Q},0)$\,, and for wave vectors ${\bf q}$
close to ${\bf Q}$ we get
$J\chi({\bf q},0)= 
   1\,/\,[({\bf q}- {\bf Q})^2 + \xi_{AF}^{-2}]$
with 
$\xi_{AF}^2= J \chi_{AF}$\,. 
Coming from high temperature or doping, an AF instability is indicated
by $(\chi_{AF})^{-1}\to 0$\,. $\chi_{AF}$ has been calculated
numerically, the resulting phase boundary is shown in Fig.\
\ref{fig-phase} as a dotted line labeled `AF'. Our result is similar
to the phase diagram obtained in Ref.\ \onlinecite{inaba96}\,. 

Apparently, at zero temperature, AF order occurs at a quite high hole
concentration $x^0_c\approx 0.22$\,, which is totally inconsistent
with experiment ($x_c\approx 0.02$). Furthermore, the study of
magnetic properties in the paramagnetic phase is bound to $x>
x_c^0$\,, i.e., the overdoped region. The high $x_c^0$ is an artifact
of the mean-field approximation. Within the gauge-field approach it
has been shown \cite{kim99} that the AF ordered state at 1/2-filling
$x=0$ is quickly removed for $x>0$\,. Furthermore it is known
\cite{leefeng88,iga92,kha93} that the interaction of spin-waves with
doped holes destroys AF order at a small $x_c$\,. In order to treat
underdoped systems we include these physics in a phenomenological
fashion. We assume a renormalization of the magnetic interaction $J_p$
in Eq.\ (\ref{eqn-sus})\,, such that $x_c^0$ is reduced to some
$x_c\approx 0.03$\,. The model is to replace
\begin{equation}  \label{eqn-alpha}
  J\to \alpha J 
    \;,\; 
  \alpha= 0.35 \leftrightarrow x_c\approx 0.03
\end{equation}
in Eq.\ (\ref{eqn-jjp})\,, $J^\perp$ stays unchanged. The actual value
of $\alpha$ is equivalent to choosing a specific critical doping
$x_c$\,. As long as $x_c$ is physically reasonable ($\le 0.05$),
results do not depend significantly on $x_c$\,. Using $J\to\alpha J$
in Eq.\ (\ref{eqn-singlrpa}) we get a new AF phase boundary, which is
shown in Fig.\ \ref{fig-phase} as the continuous line `AF'.

\begin{figure}[h]
  \centerline{
     \includegraphics[width=0.9\hsize]{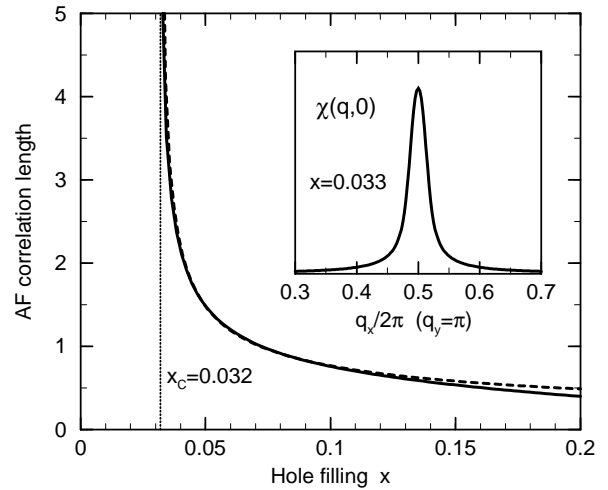} }
  \caption[\ ]{ 
  {\bf Main figure:} Antiferromagnetic (AF)
  correlation length $\xi_{AF}$ at $T=0$ in units of the lattice
  spacing. Continuous line: $\xi_{AF}(x)$ for an
  effective interaction $\alpha= 0.35$ (see text).  The vertical
  dotted line indicates the AF instability at $x_c= 0.032$\,. Dashed
  line: Function $0.2 / \sqrt{x - x_c}$\,, fitted to
  $\xi_{AF}(x)$ for $0.033\le x\le 0.1$\,. {\bf Inset:} Static
  susceptibility $\chi({\bf q},0)$ (in arbitrary units) as a wave-vector
  scan over $(\pi,\pi)$ in the 2D Brillouin zone, for $x=
  0.033$ slightly above $x_c$\,. Shown is a scan in
  $(\pi,0)$-direction, scans in other directions look similar,
  indicating a commensurate AF transition. 
  $\alpha=0.35$ and $T=0$ as above.  } 
  \label{fig-correl}
\end{figure}
To give another argument in favor of the simple interaction model,
especially the $\alpha$ being independent of doping, we consider the
correlation length $\xi_{AF}(x)$ as function of hole filling.  Fig.\
\ref{fig-correl} shows $\xi_{AF}$\,, calculated with $J\to\alpha
J$\,. It diverges at a $x_c= 0.032$ and decreases rapidly with
additional doping, following
$\xi_{AF}(x)\approx 0.2/\sqrt{x - x_c}$\,.
This behavior is consistent with neutron-scattering measurements
\cite{thu89} on LSCO and results from high temperature series for the
$t$--$J$-model \cite{singlen92}\,. The function $\sim 1/\sqrt{x}$
represents the average distance of doped holes and has been used in
\onlinecite{thu89} to interpret the data.  Finally it is noted that
the AF transition at $x_c$ occurs at the N{\'e}el wave vector ${\bf
Q}=(\pi,\pi)$\,, i.e., is commensurate. This is shown in the inset of
Fig.\ \ref{fig-correl}\,.

\section{The magnetic resonance} \label{sec-reson}
This section presents results for the magnetic response at the
antiferromagnetic (AF) wave vector ${\bf Q}= (\pi,\pi)$\,. We consider
a single CuO$_2$-layer with a nearest and next-nearest neighbor
hopping $t= 2J$ and $t'= -0.45t$\,, appropriate
\cite{schabel98ii,okand94} for YBCO\,. Effects specific to the
bi-layer structure of YBCO and BSSCO, namely the splitting of the
magnetic response into acoustic and optical modes will be discussed
separately in Section
\ref{sec-bilay}\,.

\begin{figure}[h]
  \centerline{ 
    \includegraphics[width=0.9\hsize]{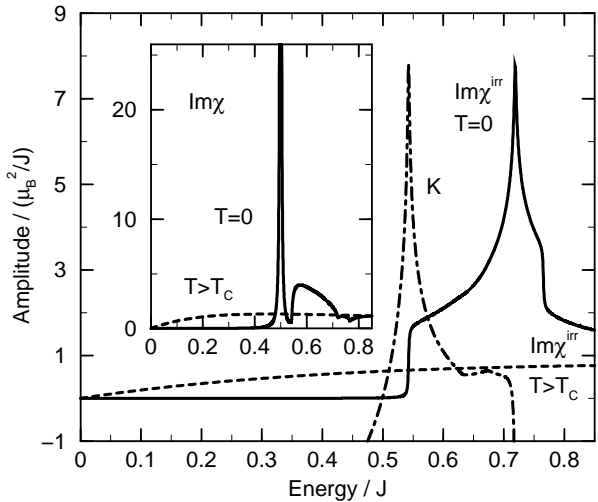} }
  \caption[\ ]{
    Magnetic response at the AF wave vector $(\pi,\pi)$ for a
  single CuO$_2$-layer near optimal doping, $x=0.12$\,. Parameters are
  $t= 2J$\,, $t'= -0.45t$\,. 
  {\bf Main figure:} 
  Imaginary part of the bubble 
  $\chi^{irr}$ in Eq.\ (\ref{eqn-singlrpa}). Shown is the
  superconductor at $T=0$ (cont.\ line)  and the normal state 
  $T=0.2J >T_c$ (dashed line). The dashed-dotted line is the inverse
  Stoner-factor $K$ (see text) for $T=0$\,, scaled $\times (-5)$\,. 
  {\bf Inset:} Imaginary part of the resulting susceptibility
  $\chi$\,, Eq.\ (\ref{eqn-singlrpa}), for $T=0$ (cont.\ line) and
  $T>T_c$ (dashed line). The sharp peak visible for $T=0$ is actually a
  $\delta$-function, broadened by a small damping used in the
    numerical calculation. 
    }
  \label{fig-pires}
\end{figure}
\subsection{Results for the AF wave vector $(\pi,\pi)$}
The dynamical susceptibility is obtained from Eq.\
(\ref{eqn-singlrpa}) with the effective interaction $J\to\alpha J$\,,
$\alpha= 0.35$ introduced in the preceding section. The integration
in Eq.\ (\ref{eqn-bubb}) is performed numerically on a $5000\times
5000$ lattice in ${\bf k}$-space, with the infinitesimal $i0_+\to i
2\Gamma$ replaced by a small finite damping $2\Gamma= 0.001J$\,.

Calculations for a fixed hole filling $x=0.12$ near optimal doping,
based on mean-field parameters from the self-consistent solution of
Eqs.\ (\ref{eqn-mfeq}) are shown in Fig.\ \ref{fig-pires}\,. In the
superconducting (SC) state at $T=0$ the imaginary part ${\rm
Im}\chi^{irr}({\bf Q},\omega)$ of the irreducible bubble is
characterized by a gap up to a threshold energy $\Omega_0\approx
0.54J$\,, with a step-like van~Hove singularity (v.H.s.) at the onset
of spectral weight at $\omega=\Omega_0$\,. A peak at $\omega\approx
2\Delta^0= 0.72J$ is remnant of the density of states of the d-wave
superconductor. The corresponding real part is shown in Fig.\
\ref{fig-pires} as the inverse Stoner-enhancement factor
$K({\bf Q},\omega)= 
   [1 - \alpha 4J\,{\rm Re}\chi^{irr}({\bf Q},\omega)]$\,. 
By virtue of the Kramers--Kroenig transformation, the step at the
threshold $\Omega_0$ in ${\rm Im}\chi^{irr}$ turns into a
log-singularity in ${\rm Re}\chi^{irr}$\,, and $K({\bf Q},\omega)$
crosses zero at an energy $\omega_{res}=0.5J <\Omega_0$ {\em within}
the gap. This leads to an undamped $\delta$-like resonance at
$\omega_{res}$ in the magnetic response ${\rm Im}\chi({\bf
Q},\omega)$\,, as is shown in the inset of Fig.\
\ref{fig-pires}\,. The position $\omega_{res}= 0.5J\approx 60$\,meV is
not too far off the $\approx 40$\,meV observed in optimally doped YBCO
and BSCCO\,.  The situation changes drastically in going to the normal
state $T>T_c\approx T_d$\,. As is seen from the dashed line in Fig.\
\ref{fig-pires} (main figure), ${\rm Im}\chi^{irr}({\bf Q},\omega)$
looses its structure, in particular the gap vanishes. The
corresponding $K$ (not shown in the figure) becomes equally
structure-less without any zero crossing, leading to a vanishing of
the resonance in ${\rm Im}\chi$ in the normal state (inset of Fig.\
\ref{fig-pires}).

\begin{figure}[h]
  \centerline{ \includegraphics[width=0.9\hsize]{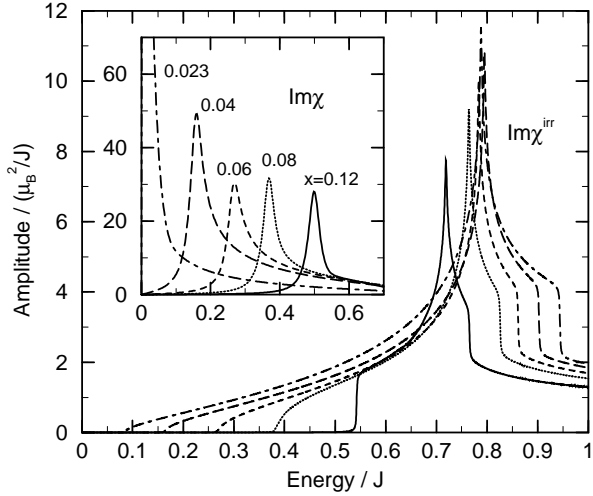} }
  \caption[\ ]{ 
    Magnetic response at wave vector $(\pi,\pi)$ as in Fig.\
  \ref{fig-pires}\,, for optimal to underdoped hole filling in the
  superconducting state $T=0$\,. 
  {\bf Main figure:} 
  Imaginary part of $\chi^{irr}$\,. Curves are identified by the
  respective onset of spectral weight (threshold), which is $0.54 J$
  for $x= 0.12$ and decreases for $x= 0.08, 0.06, 0.04, 0.023$ down to
  $0.09J$\,. 
  {\bf Inset:} Imag.\ part of the susceptibility $\chi$\,. Here the
  peaks are   broadened to an experimental resolution (FWHM) of
  $4\Gamma\approx 5$\,meV\,. 
    } 
  \label{fig-pidope}
\end{figure}
The effect of underdoping is demonstrated in Fig.\
\ref{fig-pidope}\,. For comparison with experiment ${\rm
Im}\chi$ in the inset has been broadened to an experimental resolution
(FWHM) of $4\Gamma\approx 5$\,meV through a damping $i0_+\to 2\Gamma=
0.02J$ in Eq.\ (\ref{eqn-bubb})\,.  When $x$ is reduced from $x=0.12$
the gap $\Omega_0$ in ${\rm Im}\chi^{irr}$ decreases monotonously, and
with it the resonance at $\omega_{res}\lesssim\Omega_0$ moves to lower
energies.  $\omega_{res}$ reaches zero at the AF transition, which
occurs at $x_c=0.023$ for the parameters used here ($x_c$ depends
slightly on $t'$\,, since $\alpha= 0.35$ is held fixed). Note that
$\omega_{res}$ moves opposite to the maximum gap $\Delta^0$\,. The
latter increases when $x$ decreases (see Fig.\ \ref{fig-phase}) and is
reflected in the peak in ${\rm Im}\chi^{irr}$ at higher energies
$\omega= 2\Delta^0= 0.72 \ldots 0.80$ in Fig.\
\ref{fig-pidope}\,.  The spectral weight
$W= \int\,{\rm d}\omega\,{\rm Im}\chi$ 
increases \cite{kei97} when $x$ is reduced, since the system is
shifted closer to the magnetic instability. For a quantitative
comparison of $W$ we follow a procedure applied to experimental INS
data in Ref.\ \onlinecite{kei97,fong00}\,. For $x=0.12$ the flat
normal-state spectrum is subtracted as a background from the $T=0$
curve shown in Fig.\ \ref{fig-pidope}\,, inset. Integrating only the
positive part of the resulting difference spectrum gives the weight
$\Delta W$ of the resonance compared to the normal state. We find
$\Delta W= 1.55\,\mu_B^2$\,, which agrees well with optimally doped
YBCO\,.  With reducing $x$ the resonance also develops some intrinsic
damping. The step-height at the threshold $\Omega_0$ in ${\rm
Im}\chi^{irr}$\,, which is responsible for the $\delta$-like
resonance, decreases and eventually vanishes around $x=0.09$\,.

\begin{figure}[h]
  \centerline{ 
    \includegraphics[width=0.9\hsize]{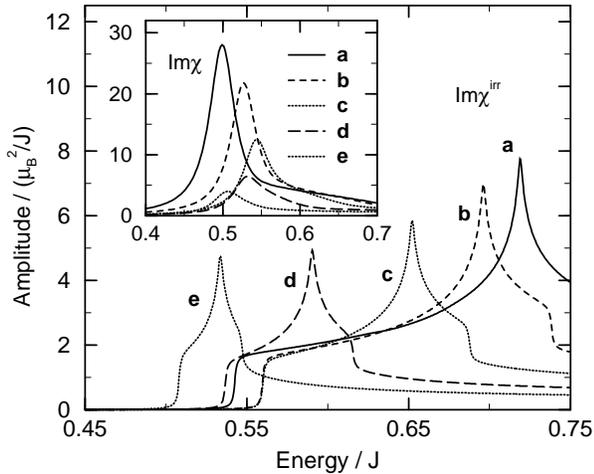} }
  \caption[\ ]{
    Magnetic response at ${\bf q}= (\pi,\pi)$ for optimal to
  overdoping at $T=0$\,. Parameters as in Fig.\ \ref{fig-pidope}\,,
  except hole filling. 
    {\bf Main figure:}
    Bubble spectrum ${\rm Im}\chi^{irr}$ for $x=0.12$
  (curve a), $0.14$ (b), $0.18$ (c), $0.24$ (d), $0.30$ (e)\,. 
    {\bf Inset:}
    Corresponding ${\rm Im}\chi$\,, broadened to $4\Gamma\approx
  5$\,meV\,. 
    }
  \label{fig-overd}
\end{figure}
The effect of overdoping is presented in Fig.\ \ref{fig-overd}\,. When
$x$ is increased from $0.12$ up to $0.3$\,, the ph-threshold
$\Omega_0$ in ${\rm Im}\chi^{irr}$ and with it the position of the
resonance first grows, but $\omega_{res}$ starts to decrease around
$x= 0.18$\,. This trend is consistent with recent INS experiments
\cite{he01} on overdoped BSCCO, where the resonance appeared at an
energy $\omega_{res}$ reduced from the optimally doped case
\cite{fong99b}\,.

\begin{figure}[h]
  \centerline{ 
    \includegraphics[width=0.9\hsize]{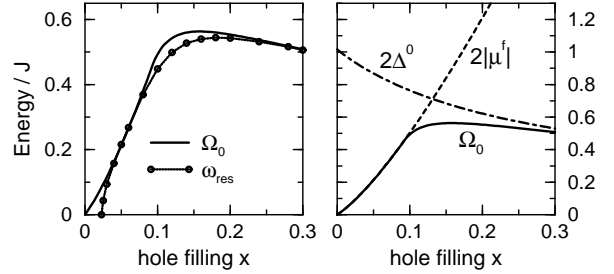} }
  \caption[\ ]{
    {\bf Left panel:}
    Particle--hole excitation threshold $\Omega_0$ and resonance
  position $\omega_{res}$ as function of hole filling in the SC state.
  $\Omega_0$ is 
  given by Eq.\ (\ref{eqn-thresh})\,, and $\omega_{res}$ is the energy
  of the peak maximum in ${\rm Im}\chi$\,, extracted from plots like
  Figs.\ \ref{fig-pidope}\,, \ref{fig-overd} (insets). 
    {\bf Right panel:}
    Comparison of $\Omega_0$\,, the chemical potential $\mu^f=
  -|\mu^f|$ of fermions\,, and the maximum gap $\Delta^0$\,. 
    }
  \label{fig-dopedisp}
\end{figure}
Fig.\ \ref{fig-dopedisp} (left) summarizes the doping dependence of
the resonance. $\omega_{res}$ is always located slightly below the
threshold $\Omega_0$ to the damping particle--hole continuum. Near the
magnetic instability
\cite{rem-piflux} at $x= x_c= 0.023$ and around optimal doping
$x\approx 0.1\ldots 0.2$\,, the resonance is a $\delta$-function, well
separated from the continuum. The $x$-dependent threshold is given by
\begin{equation}  \label{eqn-thresh}
  \Omega_0 = \left\{
    \begin{array}{lll}
      \displaystyle 
        2 |\mu^f| \sqrt{\sigma(2 - \sigma)}
        & \;\;{\rm for}\;\; &
        \sigma< 1
        \\ \\
      \displaystyle
        2 |\mu^f|
        & \;\;{\rm for}\;\; &
        \sigma\ge 1
    \end{array}
  \right.
\end{equation}
with
$\sigma= (\Delta^0)^2 \,/\, 8xt'\mu^f$\,.
For $x< \overline{x}\approx 0.09$ it is $\sigma>1$ and therefore
$\Omega_0= 2|\mu^f|$\,. That is, in the underdoped regime the
resonance energy follows the chemical potential $|\mu^f|= -\mu^f$ of
the fermions, and thus increases with hole filling. The gap
$\Delta^0$\,, on the contrary, decreases. This is illustrated in the
right panel of Fig.\ \ref{fig-dopedisp}\,. Around $x= 0.13$\,, where
$|\mu^f|=
\Delta^0$\,, a crossover into the overdoped regime occurs, where
$\Omega_0\approx 2\Delta^0$ for large $x$\,. The increase of
$\omega_{res}$ turns into a decrease. Compared to experiment
\cite{he01} the latter is too weak; this is due to $\Delta^0$ (and
$T_c= T_d$) decreasing too slowly with $x$ in the self-consistent
mean-field calculation.

It has been pointed out above that the resonance depends on the d-wave
gap to be present, i.e., the superconducting or spin-gap state, and a
sufficient next-nearest neighbor hopping $t'<0$\,. The influence of
$t'$ becomes apparent, if the bubble $\chi^{irr}$ and the resulting
magnetic spectrum ${\rm Im}\chi$ at $T=0$ are re-calculated for $t'=0$
(not shown in the figures). The response ${\rm Im}\chi$ no longer
contains a resonance-like peak, although the spectral weight is
conserved. Only for small $x\to x_c$ at the magnetic instability ${\rm
Im}\chi$ develops a Goldstone (Bragg) peak. The bubble spectrum at the
threshold $\Omega_0= 2|\mu^f|$ now follows
${\rm Im}\chi^{irr}({\bf Q},\omega)\sim
   \sqrt{\omega - \Omega_0}\Theta(\omega - \Omega_0)$
for the whole range of $x$\,, and a step-like v.H.s.\ never appears.

\subsection{Discussion}
The results presented above compare well to neutron-scattering
experiments \cite{fong00,daimook01}\,: The variation of the resonance
energy $\omega_{res}$ with hole filling $x$ is reproduced in the
underdoped \cite{dai96,fon97,kei97,mesot01pre} and the overdoped
\cite{bou96,fong99b,he01} regime. At optimal doping the resonance appears
resolution limited (as a delta function) in the SC state only
\cite{moo93,fon95}\,, and its energy and spectral weight are
comparable to the values measured in experiment
\cite{fong00}\,. In the underdoped regime it is obtained also in the
spin-gap phase \cite{dai96,fon97,mesot01pre} above $T_c$\,; the
observed line-shape (damping) is not reproduced in mean-field
theory. YBCO and BSCCO are bi-layer materials, i.e., consist of two
coupled CuO$_2$ planes per unit cell. However, for the resonance to 
emerge the bi-layer structure is not important. Rather, in optimally
and slightly underdoped systems it depends on a hole-like Fermi
surface (i.e., a sufficient $t'<0$) and a finite d-wave (spin-)
gap. As will become clear in Section \ref{sec-bilay} this conclusion
is not altered if the double-layer structure is taken into account.

The slave-particle approach reproduces quite satisfactorily the
resonance energy $\omega_{res}$ in the underdoped regime, where
$\omega_{res}$ is {\em not} connected to the maximum gap $\Delta^0$\,,
see Fig.\ \ref{fig-dopedisp}\,. $\omega_{res}$ is given by the pole of
Eq.\ (\ref{eqn-singlrpa}), i.e., the energy of the bound state in the
particle--hole channel of the fermions. In the underdoped regime
$\omega_{res}\lesssim 2|\mu^f|$ follows the chemical potential
$|\mu^f|=-\mu^f$\,. $\mu^f$ refers to quasi particles (the fermions)
which emerge from the mean-field description of the $t$--$J$-model and
are strongly renormalized. They propagate with hopping matrix-elements
$\widetilde{ t}= x t + \frac{3}{8}J\widehat{\chi}
   \approx x t + 0.15\,J$\,,
$\widetilde{ t}'= x t'$\,,
and hence Fermi velocity $\widetilde{ v}_F\approx x v_F$\,, that are
reduced from the bare parameters by the small Gutzwiller factor $x\le
0.15$\,. The latter mimics the reduced phase space due to local
correlations in the doped Mott insulator. Accordingly $|\mu^f|$ comes
out small enough, such that in the underdoped regime
$|\mu^f|<\Delta^0$\,, and $\omega_{res}\lesssim 2|\mu^f|$ is
determined by $|\mu^f|$\,, which increases with hole filling $x$\,. In
contrast, if unrenormalized quasi particles (QP) are assumed with bare
$t, t', v_F$\,, the chemical potential
$|\mu|\sim\frac{1}{x}|\mu^f|\gg\Delta^0$\,, and $\omega_{res}\lesssim
2\Delta^0$ is connected to the gap for almost all $x$\,, which
decreases with $x$\,. To achieve a satisfactory result from
unrenormalized QP the effective (residual) interaction
\cite{rem-effect} has to be made doping dependent. This is the case in
theories based on the spin-fermion model
\cite{abachu99,morr98}\,, where the coupling constant is controlled by
a magnetic correlation length, which can be chosen $x$-dependent.

\begin{figure}[h]
  \centerline{
    \includegraphics[width=0.85\hsize]{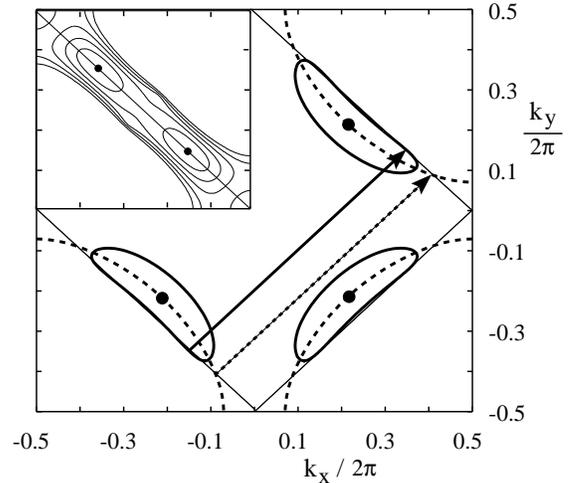} }
  \caption[\ ]{
    Origin of the threshold for particle--hole (ph) excitations with
    wave vector $(\pi,\pi)$\,, corresponding to the spectra in Fig.\
    \ref{fig-pires}\,. 
    {\bf Main figure:} 
    2D Brillouin zone (BZ) with the
    underlying normal-state Fermi surface (FS) for $t= 2J, t'= -0.45t$
    as dashed lines. The dashed arrow 
    indicates a ph-excitation with ${\bf q}=(\pi,\pi)$ at a minimal energy
    $\Omega_0=0$\,. 
    In the superconducting state the FS collapses
    to nodes (indicated as dots), and $\Omega_0>0$\,. The full arrow
    connects two constant-energy lines of quasi-particles
    $E({\bf k})= \Omega_0/2$\,. 
    {\bf Inset:}
    ph-excitation energies $\Omega({\bf q},{\bf k})$ of the
    superconductor for ${\bf q}= (\pi,\pi)$ in 
    the upper right BZ/4\,. Shown are the minima $\Omega_0$ as dots
    and the first 5 higher energies as lines. Line distance is $0.05J$\,. 
    }
  \label{fig-phexit}
\end{figure}
\subsection{Properties of the ph-threshold $\Omega_0$}
In the remainder of this section we study in some detail the origin
and qualitative properties of the threshold $\Omega_0$ in the bubble
spectrum ${\rm Im}\chi^{irr}({\bf q},\omega)$ at ${\bf
q}=(\pi,\pi)$\,. The presence of this threshold and the step-like onset
of spectral weight at $\omega=\Omega_0$ in the superconducting state
lead to the sharp resonance. We consider ${\rm Im}\chi^{irr}$ at
$T=0$\,, which reads from Eq.\ (\ref{eqn-bubb}) with
(\ref{eqn-singldisp}) for $\omega>0$
\begin{equation}  \label{eqn-imchi0}
  {\rm Im}\chi^{irr}({\bf q},\omega) \sim
    \sum_k \delta(\omega - \Omega({\bf q}, {\bf k}))\,. 
\end{equation}
BCS coherence factors have been ignored. It is determined by the
particle--hole (ph) excitation energies
$\Omega({\bf q}, {\bf k})= 
  E({\bf k}) + E({\bf k} + {\bf q})$
of fermions. The pair momentum ${\bf q}$ is set to ${\bf
Q}\equiv(\pi,\pi)$ in the following. Fig.\
\ref{fig-phexit} illustrates the situation for $x=0.12$ near optimal
doping, corresponding to the spectra shown in Fig.\
\ref{fig-pires}\,. In the normal state $T>T_c$ a ph-excitation
connects two points of the underlying Fermi surface (FS), and the
threshold
$\Omega_0= {\rm min}_k\,\Omega({\bf Q}, {\bf k})$
is 0\,. In the superconducting state a finite $\Omega_0$ is at first
sight surprising, since the d-wave SC has a finite density of
states. However, when the FS collapses to nodes, a minimum energy
$\Omega_0>0$ has to be paid for ph-excitations with wave-vector
${\bf Q}$\,. It turns out that $\Omega({\bf Q}, {\bf k})$ is
minimal, if $E({\bf k})=E({\bf k} + {\bf Q})= \Omega_0/2$\,, with the
lines of constant energy $E({\bf k})=\Omega_0/2$ touching the reduced
(magnetic) Brillouin zone at 8 points. These points are connected by
${\bf Q}$ as indicated by the full arrow in Fig.\ \ref{fig-phexit}\,.
Apparently $E({\bf k})= \Omega_0/2$ is very flat near these points,
close to a (dynamic) nesting condition. This is due to the
bandstructure of the underlying normal state, in particular the
$t'<0$\,, and the presence of the d-wave gap $\Delta({\bf k})$\,. The
resulting ph-dispersion $\Omega({\bf Q}, {\bf k})$ is shown in the
inset of Fig.\
\ref{fig-phexit}\,. It displays two minima per 1/4 Brillouin zone with
energy $\Omega({\bf Q}, {\bf k}^0)=
\Omega_0$\,, which fall on the line $k^0_x + k^0_y=
\pi$\,. The neighborhood of these minima is quite flat, and $\Omega$
can be expanded as
$\Omega({\bf Q}, {\bf k})\approx \Omega_0 + a_1 (k_1)^2 + a_2
   (k_2)^2$\,,
with $k_{1,2}= \frac{1}{\sqrt{2}}((k_x - k_x^0) \pm (k_y - k_y^0))$
and relatively small $a_1, a_2 >0$\,. Eq.\ (\ref{eqn-imchi0}) then
shows a step-like van~Hove singularity (v.H.s.) at $\Omega_0$\,,
${\rm Im}\chi^{irr}({\bf Q},\omega)\sim
   \Theta(\omega - \Omega_0)\,/\,\sqrt{a_1a_2}$\,.
The value of $\Omega_0$ is given by Eq.\ (\ref{eqn-thresh}) with
$\sigma<1$\,.

When the hole filling $x$ is reduced, the two minima in $\Omega({\bf
Q}, {\bf k})$ move closer, until they merge at $k_x^0= k_y^0 =\pi/2$
for $x= \overline{x}\approx 0.09$\,, which corresponds to $\sigma=1$
in Eq.\ (\ref{eqn-thresh})\,. The step-v.H.s.\ vanishes on the course (see
Fig.\ \ref{fig-pidope})\,, since $\Omega({\bf Q},{\bf k})$ 
becomes increasingly steep in $k_1$ direction. For
$x<\overline{x}$ we have $\sigma>1$\,, and ${\rm Im}\chi^{irr}$ may be
approximated by setting $t'=0$\,, leading to \cite{rem-kee}
${\rm Im}\chi^{irr}({\bf Q},\omega)\sim
   \sqrt{\omega - \Omega_0}\Theta(\omega - \Omega_0)$\,.
I.e., the step at the threshold $\Omega_0$ has changed into a 
$\sqrt{\omega}$ behavior.

\section{Incommensurate response} \label{sec-incomm}
The resonance at $\omega_{res}= 40$\,meV\,, as well as its relative in
underdoped samples with reduced $\omega_{res}< 40$\,meV is
characterized as a single (commensurate) peak at ${\bf q}= (\pi,\pi)$
in wave-vector space \cite{moo93,fon95,bou96,fon97}\,. Above the
resonance energy an incommensurate structure has been observed
\cite{bou97,arai99,bou00}\,, with broad maxima following a dispersion
similar to spin waves. Recently, inelastic neutron-scattering (INS)
experiments on underdoped
\cite{daimook01,mook98,arai99,dai98} and optimally 
\cite{daimook01,bou00,boukei00} doped YBCO 
revealed that also the magnetic response below the resonance position
shows incommensuration: Four distinct peaks appear at ${\bf q}=
(\pi\pm\delta,\pi)$ and $(\pi,\pi\pm\delta)$\,, which move away from
$(\pi,\pi)$ with decreasing energy, i.e., are described by some
`upside-down' dispersion \cite{arai99,bou00}\,. The
incommensurability is of ``parallel'' type, peaks are displaced in
$(\pi,0)$ and $(0,\pi)$-direction from $(\pi,\pi)$\,, similar to those
observed \cite{mas93,yam95} in LSCO\,. This is not expected in a
d-wave BCS picture \cite{tan94,lu92}\,, since at low energies the
particle--hole excitations from node to node should dominate, leading
to four peaks at ${\bf q}=(\pi\pm\delta',\pi\pm\delta')$\,. In this
section we demonstrate that a parallel incommensurability actually
occurs in a range of energies below the resonance.

\begin{figure}[h]
  \centerline{
    \rule{5mm}{0pt}\hfill 
    $(\pi,0)$-direction \hfill $(\pi,\pi)$-direction \hfill}
  \centerline{
    \includegraphics[width=0.95\hsize]{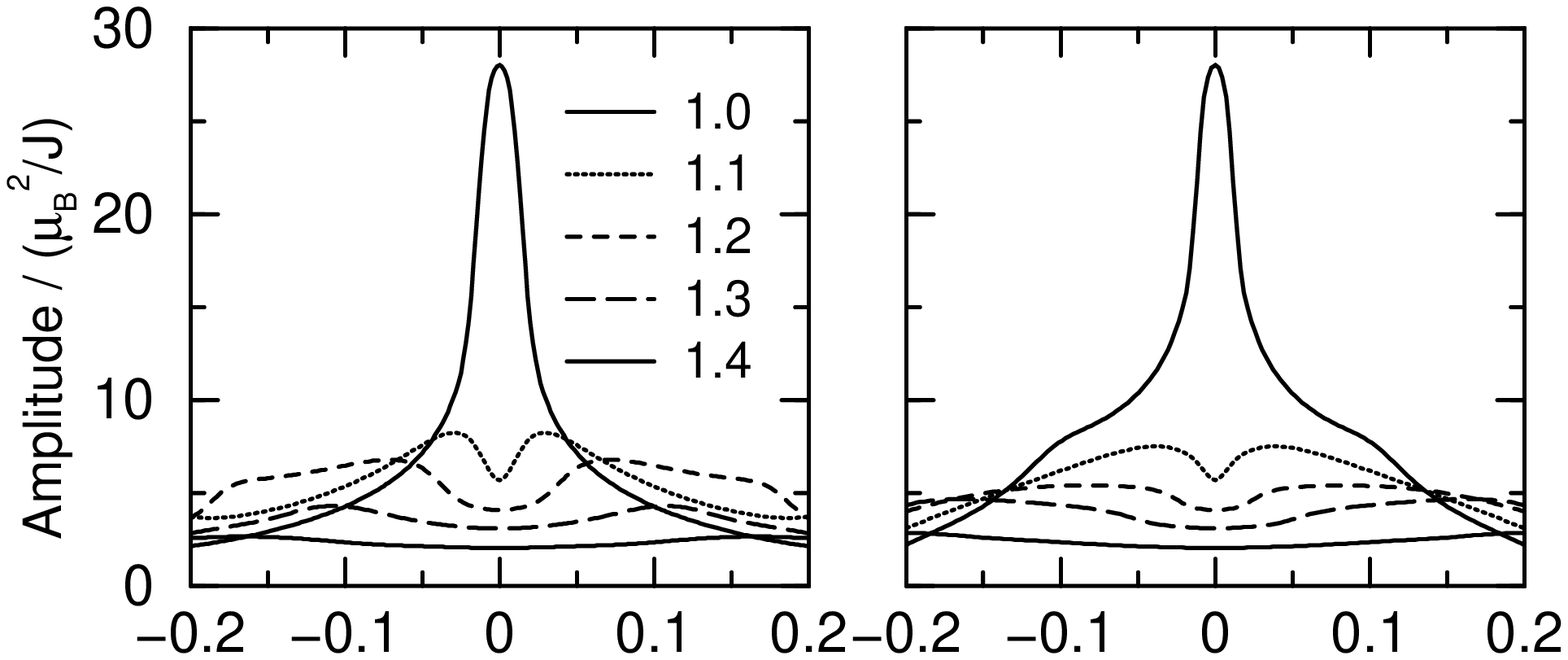} }
  \centerline{
    \includegraphics[width=0.95\hsize]{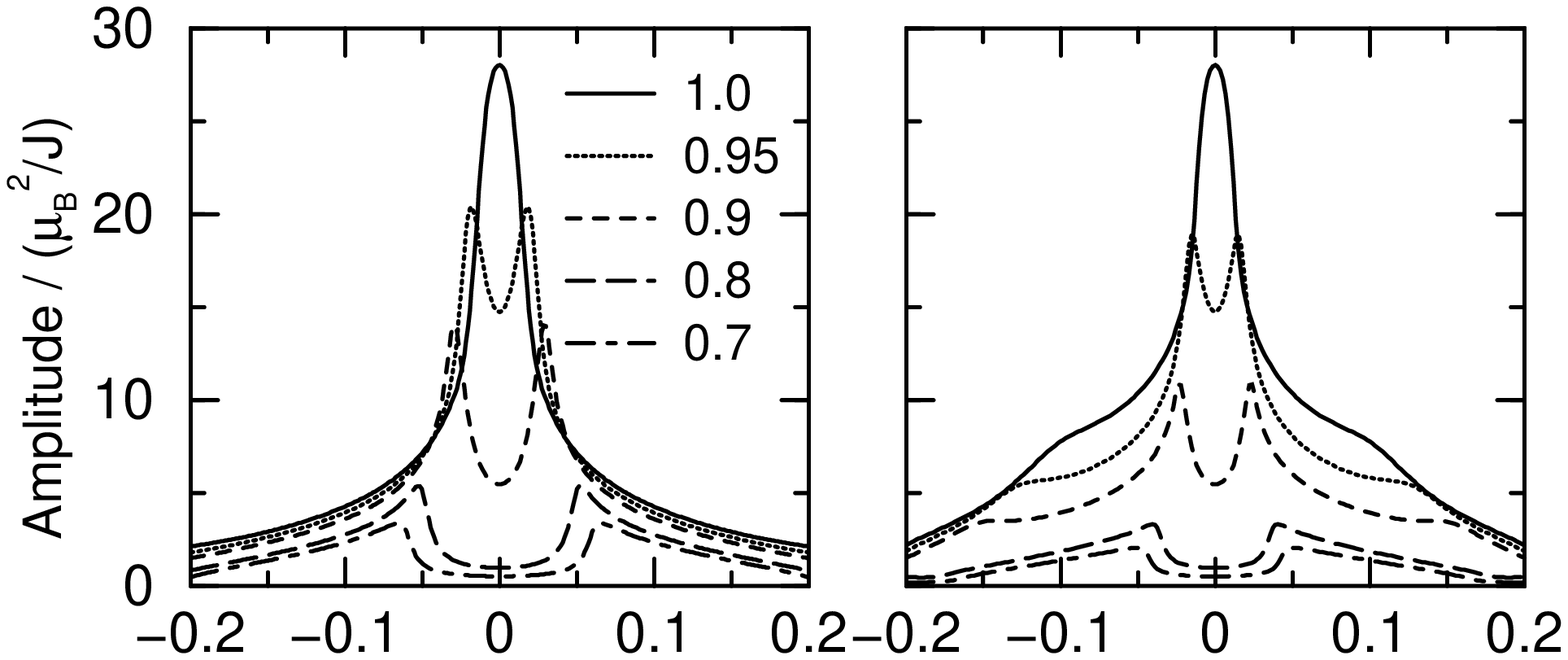} }
  \centerline{
    \includegraphics[width=0.95\hsize]{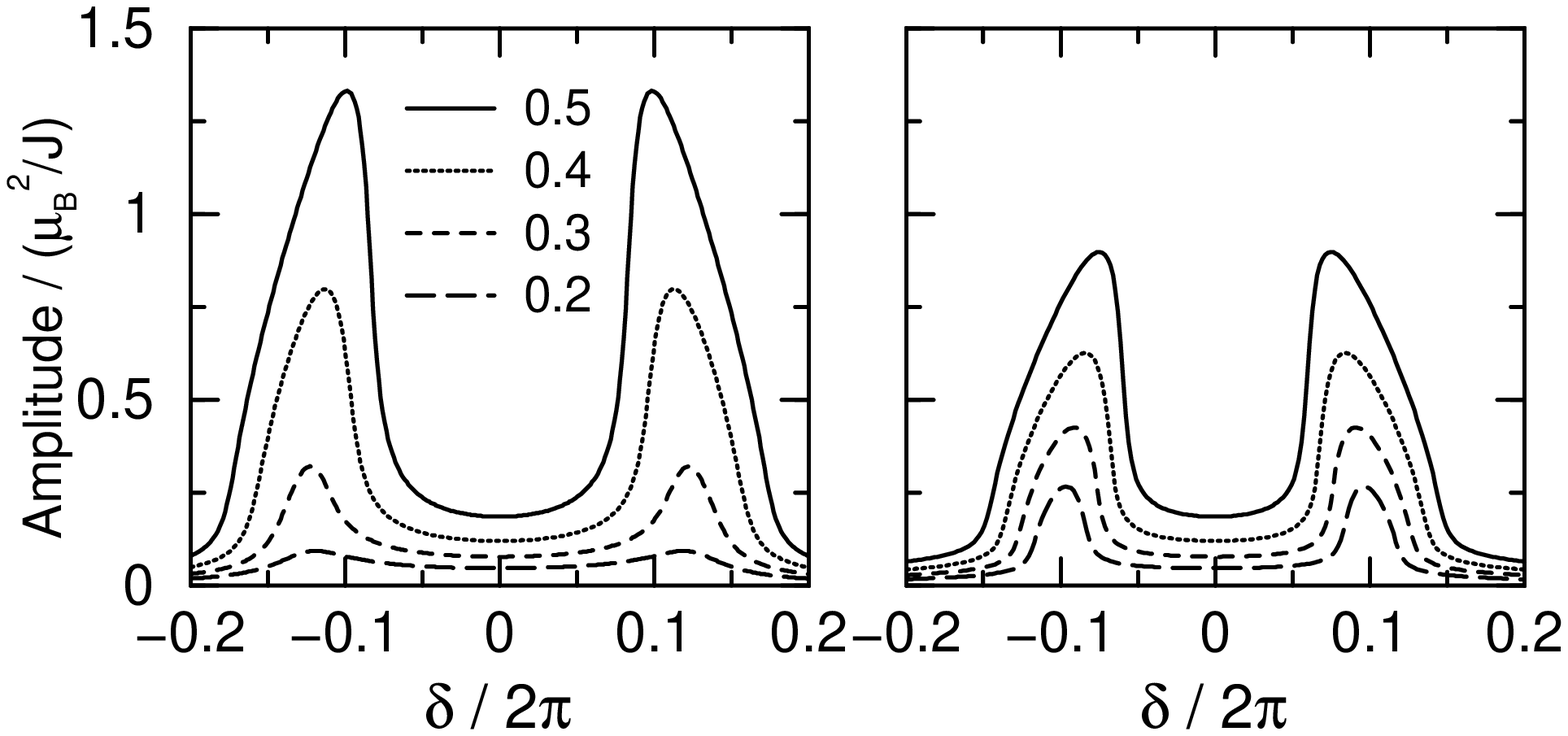} }
  \caption[\ ]{
    Wave-vector {\bf q} scans of ${\rm
  Im}\chi({\bf q},\omega)$ at fixed energy $\omega$\,, for $t=2J,
  t'=-0.45t, x= 0.12$ in the superconducting state at $T=0$\,. A
  quasi-particle damping $\Gamma= 0.01J$\,, corresponding to an
  experimental energy resolution (FWHM) of $4\Gamma\approx 5$\,meV has been
  used. {\bf q} is measured from $(\pi,\pi)$\,: $\delta_{x,y}=
  q_{x,y} - \pi$\,, $\delta= \pm\sqrt{\delta_x^2 + \delta_y^2}$\,.
  {\bf Left column:} Scans in $(\pi,0)$-direction, i.e.,
  $\delta_y= 0$\,.  {\bf Right column:} 
  $(\pi,\pi)$-direction, 
  $\delta_y= \delta_x$\,.  {\bf Top row:} Sequence of
  energies $\omega/\omega_{res}= 1.0 \ldots 1.4$ at and above the
  resonance energy $\omega_{res}= 0.5J$\,.  {\bf Middle row:} 
  $\omega/\omega_{res}= 1.0 \ldots 0.7$ at and below $\omega_{res}$\,.
  {\bf Bottom row:}
  Energies far below $\omega_{res}$\,, showing a crossover from
  parallel to diagonal incommensurability. 
  Note the different vertical scale. 
    }
  \label{fig-qscans}
\end{figure}
\subsection{Structure of the magnetic response in wave-vector space}
Fig.\ \ref{fig-qscans} presents wave-vector scans of the magnetic
susceptibility ${\rm Im}\chi({\bf q},\omega)$ for YBCO in the
superconducting state at $T=0$\,. For comparison with INS data an
experimental energy resolution $4\Gamma\approx 5$\,meV has been
simulated through a quasi-particle damping $\Gamma= 0.01J$\,. For
$\omega= \omega_{res}$ scans in parallel $(\pi,0)$ as well as diagonal
$(\pi,\pi)$-direction show a commensurate sharp peak. When energy is
increased (top panel), the resonance first evolves into a broad
incommensurate structure with maxima dispersing like spin waves
\cite{fong00,arai99}\,.  At higher energies around $2\Delta_0\approx
1.5\omega_{res}$ this turns into some featureless background. When the
energy is reduced from $\omega_{res}$ (middle panel), the peak also
splits in parallel as well as diagonal direction, suggesting a
circular structure in ${\bf q}$-space. However, the maximum intensity
is higher in $(\pi,0)$-direction, reproducing the experimental
observation.  The case $\omega/\omega_{res}\approx 0.7$ can be
compared to a study
\cite{mook98} on underdoped YBa$_2$Cu$_3$O$_{6.6}$\,: The ratio of
max.\ intensities $I_{par}/I_{diag}\approx 1.7$ from Fig.\
\ref{fig-qscans} is consistent to the $\lesssim 2.0$ we read off the
INS data in Ref.\ \onlinecite{mook98}\,. The range
$\delta=\delta_x=0.052\,\ldots\,0.065$ of the displacement of peaks
in $(\pi,0)$-direction is comparable to experimental values
\cite{rem-prldelta} reported \cite{mook98,bou00} for the same
range of energies $\omega/\omega_{res}= 0.8\,\ldots\,0.7$\,. In the
bottom panel, where the energy is reduced even further, $I_{par}$
starts to weaken relative to $I_{diag}$\,. For $\omega/\omega_{res}<
0.35$ the peaks in the diagonal $(\pi,\pi)$-direction eventually
dominate, as is expected from a d-wave superconductor in mean-field
theory at $\omega\to 0$\,. However, such a crossover from parallel to
diagonal incommensurability at low $\omega$ is not observed in
experiment. At low energies the INS data indicate a strong isotropic
suppression \cite{daimook01,bou00}\,, similar to what is seen
\cite{yam95,lake99} in LSCO\,.

In the normal state, INS on the optimally doped compound shows a broad
commensurate peak, its width depending weakly on energy
\cite{bou00}\,. From the calculation at $T>T_c\approx T_d$ we get indeed a
response almost independent of wave vector and energy, as is expected
from the absence of nesting properties of the underlying Fermi surface
(see Fig.\ \ref{fig-phexit})\,. Note that both the commensurate (at
$\omega_{res}$) and incommensurate intensity vanish in the normal
state, as is observed \cite{bou00} in INS\,. The situation is
different in underdoped YBCO\,, where the spin-gap regime is entered
as $T>T_c$\,. In mean-field theory the spin-gap phase is similar to
the SC, with the SC gap becoming the spin gap (see Section
\ref{sec-phase}). Therefore the pattern in the magnetic
response at $\omega\le\omega_{res}$ persists at $T>T_c$\,, which has
also been observed in underdoped systems
\cite{arai99,dai98}\,. Experimental line shapes at $T>T_c$ are not
reproduced in mean-field; however, we expect significant damping in
the spin-gap phase if fluctuations are included.

\subsection{The dynamic nesting effect}
An explanation of the incommensurate pattern below $\omega_{res}$ in
the superconducting state can be found in the dispersion $E({\bf k})$
of the Gutzwiller-renormalized fermions 
\cite{bri99}\,. At vanishing energy $\omega\to 0$ only particle--hole (ph)
excitations $\Omega({\bf q},{\bf k})$ with ${\bf q}$ connecting two
d-wave nodes $E({\bf k})\gtrsim 0$\,, $E({\bf k}+{\bf q})\gtrsim 0$
are possible. At finite hole doping the nodes are shifted from
$(\pm\pi/2, \pm\pi/2)$ towards the $\Gamma$-point $(0, 0)$\,; thus the
bubble spectrum ${\rm Im}\chi^{irr}$ features peaks \cite{lu92} at
${\bf q}= (\pi\pm\delta',\pi\pm\delta')$\,, diagonally displaced from
$(\pi,\pi)$\,. The curves for low $\omega/\omega_{res}= 0.2, 0.3$ in Fig.\
\ref{fig-qscans} are still dominated by this type of
ph-excitation. With increasing energy a different process gains
importance, where ${\bf q}$ connects two contours $E({\bf
k})=\omega/2$ in the Brillouin zone. Each d-wave node is surrounded by
such a `banana-shaped' contour, and in an energy range around
$\omega_{inc}\approx 0.7\omega_{res}$\,, where the parallel incommensurate
pattern in ${\rm Im}\chi({\bf q},\omega_{inc})$ is most pronounced,
these $E({\bf k})$ show almost flat pieces parallel to the magnetic
zone boundary. This gives rise to a dynamic nesting contribution
\cite{schulz90,lit93}\,, which favors peaks at ${\bf
q}=(\pi\pm\delta,\pi), (\pi,\pi\pm\delta)$\,. In particular, at
$\omega=\omega_{inc}$ a ratio of intensities $I_{par}/I_{diag}\lesssim
2$ is expected from the nesting argument, which is close to the value
drawn from the numerical calculation (Fig.\ \ref{fig-qscans}) as well
as experiment. An illustration of this effect has been given in Ref.\
\onlinecite{bri99} in Fig.\ 5\,.

In the preceding section it became apparent that the commensurate
resonance at $(\pi,\pi)$ depends on a sufficiently large n.n.n.\
hopping $t'<0$\,. With respect to the (parallel) incommensurate
pattern this is not the case: The underlying dynamic nesting
effect is a general feature of the d-wave superconductor.  This is
confirmed in a calculation of ${\bf q}$-scans for $t'= 0$\,: The
parallel incommensurability dominates for energies above the crossover
from the diagonal one and below $\approx 2|\mu^f|$\,, where ${\rm
Im}\chi$ becomes broad and commensurate.

\begin{figure}[h]
  \centerline{
    \includegraphics[width=0.95\hsize]{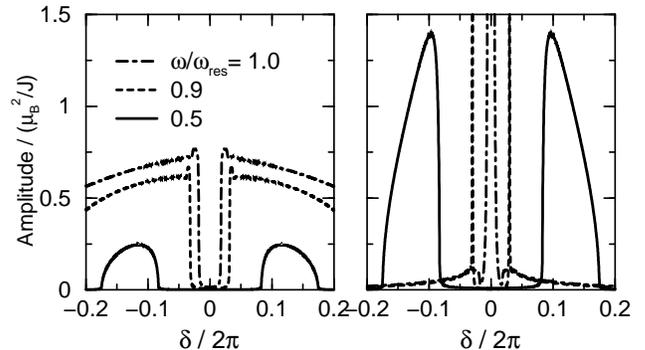}  }
  \caption[\ ]{
    Wave-vector scans in $(\pi,0)$-direction as in Fig.\
    \ref{fig-qscans}\,, left 
    column, but with $4\Gamma= 0.002J$\,. Shown are
    $\omega/\omega_{res}= 1.0, 0.9, 0.5$\,.  {\bf Left:} Bubble
    spectrum ${\rm Im}\chi^{irr}$\,.  {\bf Right:} Magn.\ response
    ${\rm Im}\chi$\,. The curves for $1.0$ and $0.9$ 
    are scaled $\times 0.01$\,; the one for $1.0$ features a 
    $\delta$-type peak, see text.
    }
  \label{fig-qscanshi}
\end{figure}
\subsection{Dispersion of the resonance}
Returning to the case $t'= -0.45t$\,, for energies increasing towards
the resonance energy $\omega_{res}$ the incommensurate pattern
eventually merges into the commensurate resonance, as is seen in Fig.\
\ref{fig-qscans}, middle row. This is due to the final-state
interaction Eq.\ (\ref{eqn-singlrpa})\,, which develops a pole at
the commensurate position $(\pi,\pi)$ for $\omega=\omega_{res}$\,. In
Fig.\ \ref{fig-qscanshi} scans similar to Fig.\ \ref{fig-qscans} are
made in $(\pi,0)$-direction, with the `experimental damping'
omitted. For $\omega/\omega_{res}=0.5$ the bubble spectrum ${\rm
Im}\chi^{irr}({\bf q},\omega)$ features two humps in
$(\pi,0)$-direction, and is zero outside these regions (left
panel). This is due to a ${\bf q}$-dependent threshold $\Omega_0({\bf
q})= {\rm min}_k\Omega({\bf q},{\bf k})$\,. With the interaction
$\alpha J$ switched on, $\chi^{irr}\to \chi$\,, the humps are merely
amplified in intensity (right panel), and the nesting argument can be
applied as above. When $\omega$ is increased, the
two humps move closer. Additionally, sharp peaks appear in ${\rm
Im}\chi$\,, which eventually merge in a single $\delta$-like resonance
at $(\pi,\pi)$ for $\omega\to \omega_{res}$\,. The incommensurate
structure still present in the bubble spectrum for
$\omega/\omega_{res}=1.0$ is completely superseded by the pole of Eq.\
(\ref{eqn-singlrpa})\,. The latter is driven by the bubble's real part
${\rm Re}\chi^{irr}$\,, which at $\omega=\omega_{res}$ is strongly
peaked at $(\pi,\pi)$ in ${\bf q}$-space, assisted by the interaction
$J({\bf q})=
   J[\cos(q_x) + \cos(q_y)]$\,,
which also favors the wave vector $(\pi,\pi)$\,. 

\begin{figure}[h]
  \centerline{
    \includegraphics[width=0.95\hsize]{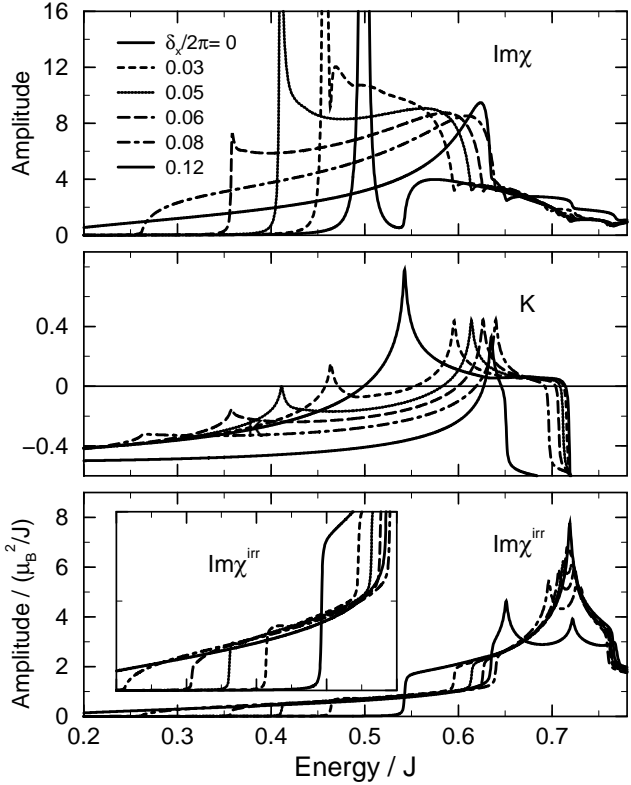}  }
  \caption[\ ]{
     Dispersion in $(\pi,0)$-direction for $x=0.12$\,. Parameters as
  in Fig.\ \ref{fig-pires}\,.  Shown 
  is a sequence of wave vectors $\delta_x=
  (0\,\ldots\,0.12)\,2\pi$\,, $\delta_y= 0$\,.  
  {\bf Bottom panel:} Bubble spectrum ${\rm Im}\chi^{irr}$\,; 
  {\em Inset:} Zoom of the threshold region. 
  {\bf Middle:} Inverse Stoner
  factor $K \times (-1)$\,.
  {\bf Top:} Resulting response ${\rm Im}\chi$ from RPA\,. 
    }
  \label{fig-escans}
\end{figure}
It is instructive to look at the commensurate--incommensurate
crossover also in energy space, using the $\omega$-scans in Fig.\
\ref{fig-escans} for several fixed wave vectors $\delta_x= q_x -
\pi$\,, $q_y= \pi$\,. At ${\bf q}= (\pi,\pi)$ the bubble spectrum
${\rm Im}\chi^{irr}({\bf q}, \omega)$ features a single step-like
onset of spectral weight at the particle--hole (ph) excitation
threshold $\Omega_0$\,; its consequences for the formation of the
neutron resonance have been discussed in Section \ref{sec-reson}
above. In Fig.\ \ref{fig-escans} (bottom) it is demonstrated how this
threshold splits into two structures as we move away from
$(\pi,\pi)$\,: the ph-threshold $\Omega_0({\bf q})$ itself, shifting
to lower energies, and a second step-like onset of additional spectral
weight at some $\Omega_2({\bf q}) \ge \Omega_0({\bf q})$\,, which shifts up
with $\delta_x$\,. The denominator (real part) of Eq.\
(\ref{eqn-singlrpa})\,,
$K({\bf q},\omega)= [ 1 + \alpha 2 J[\cos(q_x) + \cos(q_y)]
  {\rm Re}\chi^{irr}({\bf q},\omega) ]$
thus shows a splitting of the corresponding log singularity into two
(middle panel in Fig.\ \ref{fig-escans}), which in turn produces two
peaks in the magnetic response ${\rm Im}\chi({\bf q},\omega)$ (top
panel). One of these peaks disperses to lower energies and is
identified as the ${\bf q}$-dependent resonance, since it is located
below the threshold $\Omega_0({\bf q})$ and is therefore sharp. With
increasing $\delta_x$ its spectral weight is continuously reduced,
since the height of the step at $\Omega_0({\bf q})$ in ${\rm
Im}\chi^{irr}$ decreases. The other peak near $\Omega_2({\bf q})$
disperses to higher energies. It appears merely as a broad peak or
shoulder within the damping continuum. It is responsible for the
spin-wave like dispersion of broad maxima above $\omega_{res}$ in
Fig.\
\ref{fig-qscans} (top). The resonance, on the other hand, follows an
upside-down dispersion and produces the incommensurate peaks below
$\omega_{res}$ in Fig.\ \ref{fig-qscans} (middle row). 

\begin{figure}[h]
  \centerline{
    \includegraphics[width=0.8\hsize]{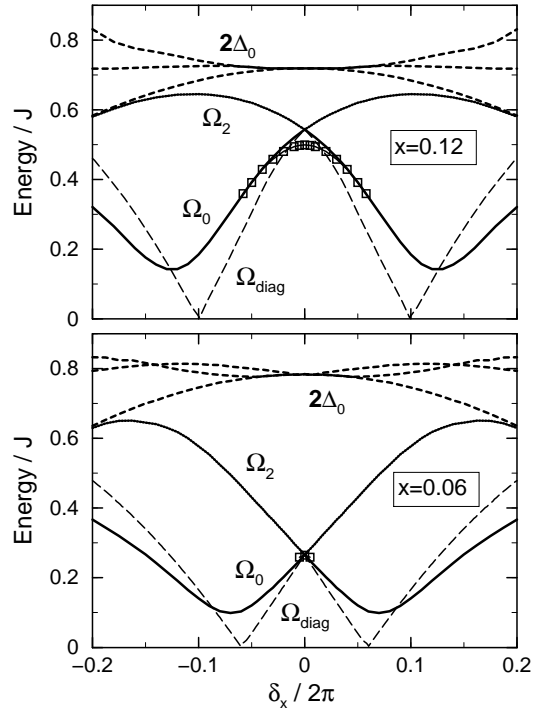} }
  \caption[\ ]{
    Dispersion of features of the bubble spectrum ${\rm Im}\chi^{irr}$
    and the spin-1 bound state (resonance) in ${\rm
    Im}\chi$\,. $\delta_x= (q_x - \pi)$\,, $q_y= \pi$\,. 
    Shown is the ph-threshold $\Omega_0$\,,
    the 2nd onset of weight $\Omega_2$\,, and the three v.H.s.\ associated
    with $2\Delta_0$\,. The position $\omega_{res}$ of the resonance
    is indicated by squares. The
    ph-threshold $\Omega_0$ is also given in $(\pi,\pi)$-direction
    as $\Omega_{diag}$\,.
    }
  \label{fig-dispers}
\end{figure}
Results for the $(\pi,0)$-direction and doping level $x=0.12$ are
summarized in Fig.\ \ref{fig-dispers} (top). It displays the
ph-threshold $\Omega_0({\bf q})$\,, the second onset of spectral
weight $\Omega_2({\bf q})$\,, and the resonance position
$\omega_{res}({\bf q})$ where it exists. For comparison the
ph-threshold $\Omega_{diag}({\bf q})$ in $(\pi,\pi)$-direction is also
shown. In contrary to $\Omega_0({\bf q})$ it has zeroes at $\delta_x=
\pm(\sqrt{2}k_F - \pi)$\,, where ${\bf q}$ connects two d-wave nodes.
In addition the energies of the van~Hove singularities (v.H.s.)
associated with $2\Delta_0$ are given in the figure. At ${\bf q}=
(\pi,\pi)$ the bubble spectrum ${\rm Im}\chi^{irr}$ (Fig.\
\ref{fig-pires}) shows a single v.H.s.\ at $\omega=2\Delta^0\approx
0.72J$\,.  From Fig.\ \ref{fig-escans} (bottom) it can be seen that
this v.H.s.\ splits into three peaks with quite flat dispersion.

The effect of strong underdoping is demonstrated in the bottom panel
of Fig.\ \ref{fig-dispers} for $x=0.06$\,: The ph-threshold $\Omega_0$
at ${\bf q}=(\pi,\pi)$ and with it the resonance position shift down
(compare also Fig.\ \ref{fig-dopedisp}). The ${\bf q}$-range, where a
sharp resonance exists, shrinks \cite{rem-qrange}\,, and the
upside-down dispersion narrows. On the other hand, the v.H.s.\ around
$2\Delta_0$ as well as the maximum of $\Omega_2$ vary only weakly with
doping ($\Delta_0$ increases slightly with underdoping).  This will
become important in the calculation of wave-vector integrated
susceptibilities in Section \ref{sec-bilay} below.

\subsection{Discussion}
The parallel type of incommensurability, i.e., a maximum intensity at
the points ${\bf q}=(\pi,\pi\pm\delta), (\pi\pm\delta,\pi)$ in the
Brillouin zone is a generic feature of the d-wave SC state. In an
energy range below $2\Delta^0$ the intensity is enhanced at these
points due to the dynamic nesting mechanism
\cite{bri99}\,. At very low energies, on the other hand, excitations
from node to node in $(\pi,\pi)$ direction dominate
\cite{lu92} and lead to a crossover to the diagonal type as $\omega\to
0$\,. It should be noted that the parallel incommensurability is not
related to ``stripes''\cite{stripe1,stripe2,stripe3}\,, i.e., we do
not consider the possibility of a combined ordering of charge and spin
into quasi one-dimensional structures. Incommensurate pattern in ${\rm
Im}\chi({\bf q},\omega)$ and their dispersion have been obtained with
similar slave-particle methods
\cite{zhalev93,kao00,ligong01pre}\,, BCS theory
\cite{salsch98,norm01,onupfe99}\,, or the FLEX approximation for
the Hubbard model \cite{takmoriya98,man01}\,. The present
slave-particle approach predicts how the dispersion of the resonance
(associated with $\Omega_0({\bf q})$ in Fig.\
\ref{fig-dispers}) and the spin-wave like dispersion of broad peaks
above it (following $\approx\Omega_2({\bf q})$) change with
underdoping. When $x$ is reduced, the peaks connected to
$\Omega_2({\bf q})$ should be observable in a wider energy range above
the resonance. Furthermore, near the bottom of $\Omega_2$ at ${\bf
q}=(\pi,\pi)$ the density of states and thus the damping is reduced,
leading to sharper peaks in the underdoped case. Experiments
\cite{fong00,bou97} actually indicate that dispersing `spin-wave'
peaks above $\omega_{res}$ are better resolved in the more underdoped
sample.

\section{Effect of the coupling in the double-layer} \label{sec-bilay}
So far we have considered a single CuO$_2$ layer as the most important
structural element of cuprate superconductors.  However, YBCO and
BSCCO contain two coupled CuO$_2$ planes per unit cell. The observed
susceptibility actually follows \cite{fong00,bou97,daimook99,hay98} 
\begin{displaymath}
  \chi({\bf q}, q_z, \omega)= 
    \chi^{+}({\bf q},\omega) \cos^2(\frac{d}{2}q_z) +
    \chi^{-}({\bf q},\omega) \sin^2(\frac{d}{2}q_z) 
    \;\,
\end{displaymath}
This form reminds of the odd $(-)$ and even $(+)$ linear combination
of spin waves in the undoped parent compound
\cite{tran89,hayaep96}\,. ${\bf q}$ is the in-plane wave vector as
before, $d$ is the distance of the 
planes within a bi-layer sandwich. We used the single-layer model as
an effective model for the odd (``acoustic'') susceptibility
$\chi^{-}({\bf q},\omega)$\,. The experimentally observed neutron
spectra in the odd mode, in particular the resonance at ${\bf q}=
(\pi,\pi)$ and its doping dependence are well reproduced by the
single-layer model. Our description of the resonance in ${\rm
Im}\chi^{-}((\pi,\pi),\omega)$ does not rely on the bi-layer structure
of the material. The important ingredient is the topology of the
underlying Fermi surface in combination with the d-wave
superconducting state. This has been discussed in detail in Section
\ref{sec-reson} above. The even (``optical'') mode spectrum ${\rm
Im}\chi^{+}((\pi,\pi),\omega)$ appears different in experiment. It
shows merely a broad peak with dim intensity \cite{fong00}\,. In this
section the calculation is extended to the bi-layer system. It is
shown that the suppression of the resonance in the even mode is mainly
a consequence of the inter-plane exchange coupling $J^\perp$\,. The
odd-mode susceptibility, on the other hand, resembles the one obtained
from the single-layer model.

The bi-layer modes have also been explored by averaging the
experimentally measured magnetic response over the in-plane Brillouin
zone \cite{fong00,bou97,daimook99,hay98}\,, 
\begin{equation}  \label{eqn-chi2d}
  {\rm Im}\chi_{2D}^{\pm}(\omega)=
    \int\!\!\!\!\int\limits_{-\pi}^\pi\!\!\frac{{\rm d}^2 q}{(2\pi)^2}
    \,{\rm Im}\chi^{\pm}({\bf q},\omega)
\end{equation}
After the ${\bf q}$-integration has been performed the odd mode
spectrum ${\rm Im}\chi_{2D}^{-}(\omega)$ is still dominated by a sharp
resonance; it occurs at the same energy $\omega_{res}\le 40$\,meV as
in ${\rm Im}\chi^{-}((\pi,\pi),\omega)$\,, but with significantly
diminished amplitude \cite{fong00}\,. In the even $(+)$ mode a second
energy scale becomes apparent. ${\rm Im}\chi_{2D}^{+}$ shows no
resonance, but a rather broad peak \cite{fong00,daimook99} or soft
onset of spectral weight \cite{bou97}\,.  The location
$\omega_{hump}\sim 80$\,meV of this `hump'-like structure is almost
independent of doping, in contrast to the strongly doping dependent
$\omega_{res}$\,. The `hump' also appears more \cite{daimook99} or less
\cite{fong00} clearly in the odd $(-)$ mode. 
$\chi_{2D}^{\pm}$ will be studied later in this section. It turns out
that particle--hole (ph) excitations across the maximum gap $\Delta^0$
lead to a hump-like peak in both modes in ${\rm Im}\chi^\pm_{2D}$\,,
at an energy $\lesssim 2\Delta^0$ almost independent of doping.

\subsection{Results for the bi-layer system}
Theoretical expressions for the susceptibility of two coupled planes
have been derived in Section \ref{sec-mean}\,. From Eqs.\
(\ref{eqn-chimeas},\ref{eqn-sus}) the mode susceptibilities are given
by
\begin{equation}  \label{eqn-birpa}
  \chi^{\pm}({\bf q},\omega)= 
    \frac{ \chi_{p}^{irr}(\omega) }
      { 1 + \widetilde{J}^{\pm}({\bf q})\chi_{p}^{irr}(\omega) }
\end{equation}
in units of $(g\mu_B)^2$\,. $\chi_{p}^{irr}$ is obtained from Eq.\
(\ref{eqn-bubb}) with $p\equiv({\bf q},p_z)$ and $p_z= \{0, \pi\}$ for
the modes $\{+, -\}$\,. The $\chi^{\pm}$ differ in their respective
irreducible particle--hole bubble $\chi_{p}^{irr}$ and the effective
interaction Eq.\ (\ref{eqn-alpha})\,,
\begin{displaymath}
  \widetilde{J}^{\pm}({\bf q})= 
    \alpha J({\bf q}) \pm J^\perp
    \;,\;
    J({\bf q})= 2J[ \cos(q_x) + \cos(q_y) ]
\end{displaymath}
For the in-plane parameters we take $\alpha= 0.35, t=2J, t'=-0.45t$ as
before, and for the coupling of the two CuO$_2$-planes within a
bi-layer we chose an antiferromagnetic exchange $J^\perp= 0.2J$ and an
inter-plane hopping \cite{chasud93,okand94}
\begin{displaymath}
  t^\perp({\bf q})= 
    2 t^\perp[ \cos(q_x) - \cos(q_y) ]^2 + t^\perp_0
\end{displaymath}
with $t^\perp= 0.1t$ and $t^\perp_0= 0$\,. 

We assume an in-plane superconducting order parameter $\Delta^0$ with
equal amplitude and phase in both layers. The self-consistent solution
of the mean-field equations (\ref{eqn-mfeq}) then leads to a vanishing
inter-plane gap $\Delta^{\perp 0}=0$ (Ref.\ Eqs.\
(\ref{eqn-gapfun},\ref{eqn-pair})) and a very small uRVB amplitude
$\widehat{\chi}^\perp\approx 0.03$\,, which has been defined in Eq.\
(\ref{eqn-hopamp})\,. Therefore the influence of $t^\perp, J^\perp$ on
the fermions that constitute $\chi_{p}^{irr}$ is merely a small
splitting of the bandstructure into bonding and anti-bonding bands
through the effective inter-plane hopping $\widetilde{ t}^\perp({\bf
k})\approx x t^\perp({\bf k})$\,.

\begin{figure}[h]
  \centerline{
    \includegraphics[width=0.95\hsize]{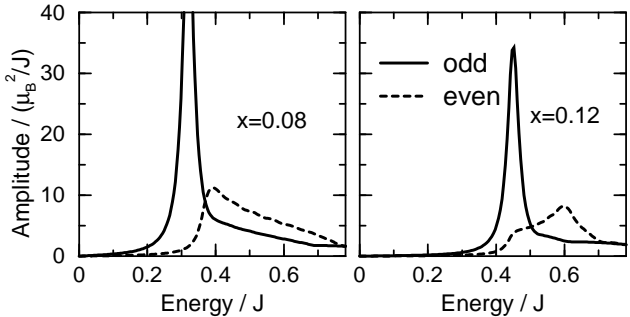}  }
  \caption[\ ]{
    Odd- and even-mode susceptibility (imag.\ part) of a bi-layer
  system, for fixed in-plane wave vector ${\bf q}=
  (\pi,\pi)$\,. Parameters are $t=2J$, $t'=-0.45t$,
  $\alpha= 0.35$, $T=0$, $4\Gamma= 0.04J\approx 5$\,meV 
  as in the preceding sections, and
  $J_\perp=0.2J$, $t_\perp=0.1t$\,. 
    }
  \label{fig-bireson}
\end{figure}
Results for fixed in-plane wave vector ${\bf q}=(\pi,\pi)$ are
presented in Fig.\ \ref{fig-bireson}\,, with `experimental' energy
resolution $4\Gamma= 5$\,meV\,. A resonance appears in the odd mode
susceptibility, which varies with doping as in the single-layer
case. The even mode, on the other hand, shows a broad peak with much
reduced intensity. This is mainly due to the mode-dependent interaction
$\displaystyle
  |\widetilde{ J}^+(\pi,\pi)| < |\widetilde{ J}^-(\pi,\pi)|$
in Eq.\ (\ref{eqn-birpa})\,, which shifts the pole in the even $(+)$
mode into the damping continuum. The damping effect is supported by
the above-mentioned splitting of fermion bands. It should be
emphasized that the resonance in the even mode is not totally
suppressed, but shifted and strongly damped. In experiment
\cite{fong00} a similar observation has been made.

\begin{figure}[h]
  \centerline{
    \includegraphics[width=0.95\hsize]{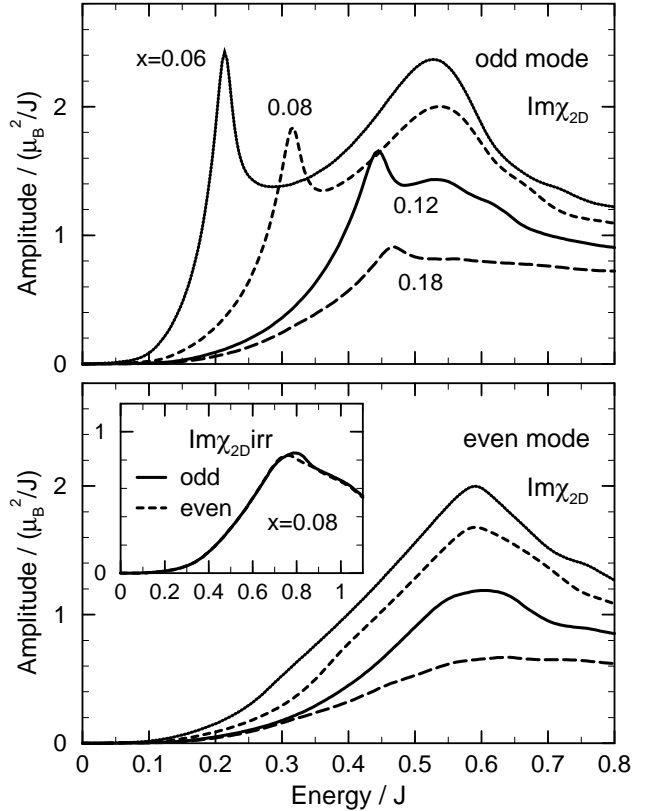}  }
  \caption[\ ]{
    Wave-vector ${\bf q}$ integrated 
    odd- and even-mode susceptibilities ${\rm Im}\chi_{2D}$ from Eq.\
  (\ref{eqn-2dmod})   for hole filling $x= 0.06\ldots 0.18$\,.
  Parameters as in Fig.\ \ref{fig-bireson}\,. 
  {\bf Inset:} 
  ${\bf q}$-integrated bubble spectrum ${\rm Im}\chi_{2D}^{irr}$ for
  $x=0.08$\,. The maximum is located close to $2\Delta^0= 0.78J$\,. 
    }
  \label{fig-chi2d}
\end{figure}
A new feature appears if we look at the wave-vector integrated
susceptibility Eq.\ (\ref{eqn-chi2d})\,, which is shown in Fig.\
\ref{fig-chi2d}\,. Similar to the case of fixed ${\bf q}=(\pi,\pi)$ a
resonance appears only in the odd mode. It appears at the same
position as in ${\rm Im}\chi^-(\pi,\pi)$ with the same strong doping
dependence. Additionally, both modes ${\rm Im}\chi^\pm_{2D}(\omega)$
show a broad peak (`hump') at an energy somewhat below $2\Delta^0$\,,
almost independent of doping ($2\Delta^0\approx 0.78J$ for
$x=0.08$)\,. For an explanation of this `hump' we first go back to the
single-layer case: The spectrum ${\rm Im}\chi^{irr}({\bf q},\omega)$
of the irreducible ph-bubble shown in Fig.\ \ref{fig-escans} (bottom)
is dominated by peaks around $2\Delta^0$\,. These van~Hove
singularities (v.H.s.)  follow the quite flat ${\bf q}$-space
dispersion shown in Fig.\
\ref{fig-dispers}\,. When the wave-vector is integrated over in
${\rm Im}\chi_{2D}^{irr}(\omega)=
    \int\!\!\!\!\int\limits_{-\pi}^\pi\!\!\frac{{\rm d}^2 q}{(2\pi)^2}
    \,{\rm Im}\chi^{irr}({\bf q},\omega)$\,,
the v.H.s.\ contribute a large density of states, leading to a broad
peak with maximum at $\omega= 2\Delta^0$\,. This argument extends to
the bi-layer system: The hump appears almost identically in both modes
of the bubble ${\rm Im}\chi^{\pm\,irr}_{2D}(\omega)$\,, which is shown
in the inset of Fig.\ \ref{fig-chi2d}\,. Its position follows
$2\Delta^0$ to slightly higher energies for reduced doping level
$x$\,.

When the final-state interaction Eq.\ (\ref{eqn-birpa}) is switched on
in the odd $(-)$ mode, the resonance appears (see Fig.\
\ref{fig-chi2d} top). Since spectral weight is 
shifted to lower energies, the hump is relocated to an $x$-independent
position below $2\Delta^0$\,. The even $(+)$ mode (Fig.\
\ref{fig-chi2d} bottom) experiences a weaker
renormalization through Eq.\ (\ref{eqn-birpa}), no resonance is
formed, and its hump is relocated less strongly. Recent FLEX
calculations for a 3-band single-layer Hubbard model in the overdoped
regime \cite{takmoriya98} give results for ${\rm Im}\chi_{2D}(\omega)$
comparable to our odd mode susceptibility.

The intensity of the resonance in ${\rm Im}\chi^-_{2D}(\omega)$ is
much reduced from its value in ${\rm Im}\chi^-((\pi,\pi),\omega)$
measured at fixed wave vector. Therefore in ${\rm Im}\chi_{2D}$ the
resonance at $\omega_{res}$ and the excitations across the maximum gap
at $\omega_{hump}\lesssim 2\Delta^0$ are of comparable intensity and
can both be observed experimentally.  This is due to the fact that the
latter occupy a large part of the in-plane Brillouin zone (BZ), while
the resonance is just a narrow peak in ${\bf q}$-space. Despite its
large amplitude, the resonance contributes only little to Eq.\
(\ref{eqn-chi2d})\,.  For the actual computation of ${\rm
Im}\chi^\pm_{2D}$ we used wave-vector scans
\cite{rem-model} in $(\pi,0)$-direction like those shown in Fig.\
\ref{fig-qscans} (extended to the whole BZ) and assumed
\cite{rem-isotrop} a susceptibility isotropic around $(\pi,\pi)$\,, i.e.,
\begin{equation}  \label{eqn-2dmod}
  \chi^\pm_{2D}(\omega)= \frac{1}{2\pi}\int_0^\pi k\,{\rm
    d}k\,\chi^\pm({\bf q},\omega) \;,\; {\bf q}= (\pi + k, \pi)
\end{equation}
The resonance is actually so sharp in ${\bf q}$-space (see Fig.\
\ref{fig-qscans}) that it does not become visible in the resulting ${\rm
Im}\chi^-_{2D}(\omega)$\,. For each $\omega$ the respective ${\bf
q}$-scan has therefore been convoluted with a Gaussian of
FWHM$=0.25$\,r.l.u.$=0.5\pi$\,, in order to simulate a finite
instrumental wave-vector resolution. Application of Eq.\
(\ref{eqn-2dmod}) then leads to the curves shown in Fig.\
\ref{fig-chi2d}\,. 

\subsection{Comparison to experiment}
Two experimental groups studied the wave-vector integrated magnetic
response ${\rm Im}\chi^\pm_{2D}$ in underdoped YBCO\,.  Refs.\
\onlinecite{daimook99,hay98} reported a lineshape for YBCO$_{6.6}$ which
agrees quite well with the theoretical result Fig.\ \ref{fig-chi2d}
for $x\le 0.08$\,. A `hump' in ${\rm Im}\chi^+_{2D}$ (even) appears at
$\approx 100$\,meV\,, ${\rm Im}\chi^-_{2D}$ (odd) shows a similar
structure at a somewhat lower energy $\approx 90$\,meV\,. The
well-known resonance appears only in ${\rm Im}\chi^-_{2D}$\,, at
$34$\,meV\,. In Refs.\ \onlinecite{fong00,bou97} two underdoped
samples YBCO$_{6.7}$ and YBCO$_{6.5}$ have been studied. In the even
(``optical'') mode of YBCO$_{6.7}$ a hump appears around $70$\,meV\,,
whereas the odd (``acoustic'') mode shows a weak hump-like structure
at $\approx 55$\,meV\,, separated from the resonance at
$33$\,meV\,. In the more underdoped sample YBCO$_{6.5}$ these features
tend to move up in energy, while the resonance in ${\rm
Im}\chi^-_{2D}$ shifts down to $25$\,meV\,.

Although the detailed experimental lineshapes are not unique
\cite{rem-backg}\,, the qualitative features of our calculation are
found in the neutron-scattering spectra. In particular we obtain the
different dependence on doping level of the resonance at
$\omega_{res}$ in the odd mode and the hump-like feature at
$\omega^\pm_{hump}$ in both modes. Also is $\omega^-_{hump}$ of the
odd mode lower than the $\omega^+_{hump}$ of the even mode. Theory and
experiments can also partly be compared quantitatively.  The measured
neutron-scattering intensities \cite{fong00,daimook99} are of the same
order as the theoretical ones in Fig.\ \ref{fig-chi2d} (using
$J=120$\,meV\,, i.e., $1\mu_B^2/J= 8.3\mu_B^2/$eV). The maximum of the
hump in the even, odd mode in Fig.\
\ref{fig-chi2d} occurs at $\omega^{+,-}\approx 0.6J, 0.53J= 72$\,meV,
$64$\,meV\,, in good agreement with the measurements \cite{fong00} on
YBCO$_{6.7}$ at low temperature.
Note that the maximum gap $\Delta^0= 30-45$\,meV \cite{rem-gap} is
consistent to the value from the mean-field calculation (see Section
\ref{sec-mean}), $2\Delta^0$ is the upper limit for the hump position in
Fig.\ \ref{fig-chi2d}\,.

\subsection{Discussion}
The wave-vector integrated magnetic response in underdoped systems is
characterized by two energy scales with opposite dependence on the
doping level. The first is the position $\omega_{res}$ of the
``41\,meV resonance'', which appears in the odd $(-)$ mode for fixed
wave vector ${\bf q}=(\pi,\pi)$ as well as in the ${\bf q}$-integrated
susceptibility. It moves down in energy when doping $x$ is reduced and
becomes a Bragg peak at the transition to the AF ordered state at $x=
x_c$\,. The second is essentially the maximum gap $\Delta^0$\,, which
increases with reduced $x$\,. It determines the position
$\omega_{hump}$ of the additional broad peak (`hump'). The latter
appears in both bi-layer modes, but only if the in-plane wave-vector
is integrated over. The hump is caused by particle--hole excitations
across $2\Delta^0$\,, and is pulled down somewhat by the final-state
interaction Eq.\ (\ref{eqn-birpa})\,. It should be noted that it is
very robust against a variation of the next-nearest-neighbor hopping
$t'$\,, i.e., the topology of the fermion's band structure and Fermi
surface. Whereas the resonance vanishes for $t'=0$ the hump remains
almost unaffected. The mechanism is very much different from the
optical spin waves that appear in the undoped $x=0$ bi-layer system if
a finite N{\'e}el order parameter is taken into account. Therefore the
appealing similarity of the hump- (or threshold-like) feature in the
superconducting YBCO samples and the optical spin-wave gap seen in the
undoped parent compound \cite{hayaep96} is accidental.

\section{Summary and outlook} \label{sec-concl}
This paper presented a theory for the magnetic excitation spectrum of
YBa$_2$Cu$_3$O$_{6+y}$ (YBCO) and Bi$_2$Sr$_2$CaCu$_2$O$_{8+\delta}$
(BSCCO) superconductors. We considered the so-called ``41\,meV
resonance'' at fixed in-plane wave vector ${\bf q}=(\pi,\pi)$\,, the
magnetic response in ${\bf q}$-space, the pecularities due to the
bi-layer structure of YBCO and BSCCO, and the local (${\bf
q}$-integrated) susceptibility. Most of the results are in good
agreement with the neutron-scattering experiments. The resonance is
obtained as a collective spin-1 excitation in the superconducting and
spin-gap states (the latter corresponding to the pseudo-gap regime of
cuprates).  Its energy scale and spectral weight as function of the
doping level $x$ at low temperature are satisfactorily reproduced. The
absence of damping in optimally doped systems is caused by the d-wave
superconducting gap in connection with the hole-type topology of the
underlying Fermi surface. The bi-layer structure is not necessary for
the resonance to form in the odd-mode (acoustic) susceptibility. The
mere effect of the finite inter-layer coupling $J^\perp$ is an almost
suppression of the resonance in the even (optical) mode. The observed
pattern of incommensurate peaks in ${\bf q}$-space has been traced
back to a dynamic nesting effect of the d-wave superconductor, and the
peak's dispersion has been derived for optimally and underdoped
systems. Besides the resonance a second, hump-like feature appears in
the wave-vector integrated magnetic spectrum. It is caused by
particle--hole excitations across the maximum gap $\Delta^0$ that
occupy a large area in the 2D Brillouin zone. Their energy
$\omega_{hump}\lesssim 2\Delta^0$ is almost independent of hole
filling, in strong contrast to the resonance position
$\omega_{res}$\,.  Also is this hump insensitive to the Fermi-surface
topology.

A salient property of the resonance at $(\pi,\pi)$ is the variation of
its energy $\omega_{res}$ with hole filling $x$\,. The $t$--$J$-model,
i.e., the doped Mott insulator naturally provides the energy scales
for the resonance in the underdoped and overdoped regimes: The
mean-field theory describes magnetic excitations in terms of quasi
particles (QP) (the fermions) with a Fermi velocity $\widetilde{
v}_F\approx (x + 0.15J/t)\,v_F$ reduced from the bare parameter. Hence
in underdoped systems the QP's chemical potential is smaller than the
gap, $|\mu^f|<\Delta^0$\,, and determines the scale for the resonance
energy as $\omega_{res}\lesssim 2|\mu^f|$ (see Fig.\
\ref{fig-dopedisp}). Thus $\omega_{res}$ is found to increase with
hole filling $x$\,, in accordance with experiment. In the overdoped
regime, on the other hand, we have $|\mu^f|>>\Delta^0$\,, and the
resonance $\omega_{res}\lesssim 2\Delta^0$ is connected to the
gap which decreases with $x$\,.

The mean-field theory in its present form overestimates the
antiferromagnetic (AF) state in the phase diagram. Therefore we had to
introduce the phenomenological parameter $\alpha$\,, which reduces the
interaction $J\to \alpha J$ in the spin-flip particle--hole channel
Eq.\ (\ref{eqn-singlrpa}) of the quasi particles. The present study
shows that already the simplest model $\alpha({\bf
q},\omega,x)=\alpha$ leads to consistent results for
neutron-scattering spectra and magnetic correlation length
$\xi_{AF}(x)$ in the relevant range of doping, energy, and wave
vector. With respect to the doping dependence of $\omega_{res}$ and
$\xi_{AF}$ this is due to the above-mentioned renormalization of the
QP\,.  In the half-filled case $x=0$\,, which we did not consider
here, the mean-field theory delivers a N{\'e}el state with the correct
spin-wave velocity only if $J$ is kept unrenormalized, i.e.,
$\alpha(x=0)=1$\,. With doping the AF state is destroyed by the
propagation of holes in the spin background
\cite{leefeng88,iga92,kha93}\,. We expect that a refined theory, where
these processes are included as corrections to mean field, yields an
$\alpha(x)$ which descreases quickly in the AF region $x<x_c$ and then
levels off in the paramagnetic (superconducting) phase $x>x_c$\,. This
is subject to future work.

We thank P.~W{\"o}lfle for useful comments on the manuscript. This
work has been supported by the NSF under MRSEC Program No.\
DMR\,98-08941 and the Deutsche Forschungsgemeinschaft through
SFB\,195\,.

\sloppy


\begin{thebibliography}{100}
\bibitem{rem-gap}
ARPES measurements on untwinned YBCO$_{6.995}$ reported in Ref.\
  \onlinecite{shen01} are consistent with maximum gap values $\Delta^0=29$\,meV
  and $44$\,meV at $(\pi,0)$ (X-point) and $(0,\pi)$ (Y-point).
\bibitem{rem-prlchi}
Note that in Ref.\ \onlinecite{bri99} slightly different units have been used:
  $\chi^{meas}= (g\mu_B)^2\chi/2$\,, with an irred.\ part $\chi^0=
  2\chi^{irr}$\,.
\bibitem{rem-piflux}
For hole filling below $x_c= 0.023$ the mean-field theory and therefore Fig.\
  \ref{fig-dopedisp} is not valid, since a finite magnetic order parameter is
  not contained. The limit $x\to 0$ here reproduces the $\pi$-flux phase
  \cite{affmar88}\,, with nodes at $(\pm\pi/2, \pm\pi/2)$ in the fermion
  dispersion, leading to $\Omega_0\to 0$\,.
\bibitem{rem-effect}
In an approach based on quasi particles the susceptibility is given by the same
  RPA formula Eq.\ (\ref{eqn-singlrpa}), with the (assumed) effective
  interaction replacing $2J[\ldots]$\,.
\bibitem{rem-prldelta}
Note that in Ref.\ \onlinecite{bri99} and some experimental work $\delta\equiv
  2\delta_x$ is used.
\bibitem{rem-model}
This procedure is similar to the one used in the experimental study Ref.\
  \onlinecite{fong00}\,. Use of wave-vector scans in directions other that
  $(\pi,0)$ causes only minor changes in our results.
\bibitem{rem-isotrop}
The resonance is in fact isotropic around $(\pi,\pi)$\,, for the excitations
  around $2\Delta^0$ this is an acceptable model.
\bibitem{rem-backg}
We suggest that this is (at least partly) due to different procedures used for
  subtracting an energy dependent background. In the theoretical calculation a
  subtraction of a background from the wave-vector scans leads to a reduction
  of intensity and change in shape of the `hump', while the resonance is not
  affected. The background is assumed constant throughout the Brillouin zone
  and is determined for each energy as the minimum of the wave-vector scan in
  $(\pi,0)$ direction. Fig.\ \ref{fig-chi2d} is computed from the unsubtracted
  (bare) scans as decribed in the text.
\end{thebibliography}
\end{document}